\documentclass[twocolumn]{jpsj2} 
\newcommand{\nn}{\nonumber}

\title{
Imbalanced Superfluid Phase of a Trapped Fermi Gas \\ in the BCS-BEC Crossover Regime
}

\author{Takeshi \textsc{MIZUSHIMA}\thanks{E-mail address: mizushima@mp.okayama-u.ac.jp}, 
Masanori \textsc{ICHIOKA}, and Kazushige \textsc{MACHIDA} }

\inst{Department of Physics, Okayama University, Okayama 700-8530, Japan}

\recdate{\today}

\abst{
We theoretically investigate the ground state of trapped neutral fermions with population imbalance in the BCS-BEC crossover regime. On the basis of the single-channel Hamiltonian, we perform full numerical calculations of the Bogoliubov-de Gennes equation coupled with the regularized gap and number equations. The zero-temperature phase diagram in the crossover regime is presented, where the Fulde-Ferrell-Larkin-Ovchinnikov (FFLO) pairing state governs the weak-coupling BCS region of a resonance. It is found that the FFLO oscillation vanishes in the BEC side, in which the system under population imbalance turns into a phase separation (PS) between locally binding superfluid and fully polarized spin domains. We also demonstrate numerical calculations with a large particle number $\mathcal{O}(10^5)$, comparable to that observed in recent experiments. The resulting density profile on a resonance yields the PS, which is in good agreement with the recent experiments, while the FFLO modulation exists in the pairing field. It is also proposed that the most favorable location for the detection of the FFLO oscillation is in the vicinity of the critical population imbalance in the weak coupling BCS regime, where the oscillation periodicity becomes much larger than the interparticle spacing. Finally, we analyze the radio-frequency (RF) spectroscopy in the imbalanced system. The clear difference in the RF spectroscopy between BCS and BEC sides reveals the structure of the pairing field and local ``magnetization''. 
}

\kword{quantum atomic gas, FFLO state, BCS-BEC crossover, imbalanced superfluid, Bogoliubov-de Gennes equation, phase diagram, radio-frequency spectroscopy
}

\begin{document}
\maketitle

\section{Introduction}

There is increasing interest in the investigation of neutral Fermi systems with mismatched Fermi surfaces.~\cite{sheehyPR} The robustness of superfluidity against the ``paramagnetic'' depairing is a longstanding fundamental issue that has captured the attention of researchers in various fields, ranging from condensed matter to color superconductivity in dense quark matter.~\cite{casalbuoni} Recently, superfluid (SF) phases under population imbalance have been realized in a trapped Fermi gas,~\cite{rice1,rice2,mit1,mit2,mit3,mit4} accompanied with the manipulation of the $s$-wave scattering length $a$ by Feshbach resonance. 

For the achievement of superfluidity in neutral atom systems, the Feshbach resonance is a key factor. Applying an external magnetic field enables one to control the relative energy between the two channels of the scattering process: the open (scattering) and closed (bound) channels. Two fermions distributed in hyperfine spin states labeled as $\sigma \!=\! \uparrow,\downarrow$ form the Cooper pair via the weak attractive interaction in the system with a negative $a$. In contrast, for the positive $a$, the energy of the closed channel, i.e., the binding energy, is characterized as $E_b \!=\! -1/Ma^2$ for mass $M$, which leads to the stable formation of tightly bound molecular bosons. The composite bosons with a long lifetime turn into the Bose-Einstein condensed phase below the critical temperature. Hence, the manipulation of interatomic interaction continuously changes the superfluidity from the fermionic Bardeen-Cooper-Schrieffer (BCS) type to molecular Bose-Einstein condensation (BEC) through the unitary limit on resonance, i.e., the BCS-BEC crossover.~\cite{review}

In actual experiments on neutral atoms, the total particle number $N$ is conserved. Applying a radio-frequency (RF) field, one can control the population difference between two hyperfine spin states, called the population imbalance,
\begin{eqnarray}
P \equiv \frac{N_{\uparrow}-N_{\downarrow}}{N_{\uparrow}+N_{\downarrow}},
\end{eqnarray}
where $N_{\sigma}$ is the number of spin $\sigma$ species. The negligible contribution of the dipole-dipole interaction leads to the conservation of population imbalance $P$ throughout the typical experimental time scale. It is known that in the presence of population imbalance, i.e., $P\!\neq\! 0$, the uniform superfluid state cannot be thermodynamically stable. Various candidates for the pairing state when $P\!\neq\! 0$ situation have been proposed, including the Fulde-Ferrell-Larkin-Ovchinnikov (FFLO) modulated pairing state,~\cite{ff,lo} the BCS-normal phase separation (PS),~\cite{bedaque} the breached pairing or Sarma state,~\cite{wilczek} the deformed Fermi surface superfluid (DFS),~\cite{muther,sedrakian} and the $p$-wave pairing state.~\cite{mur-petit,bulgacPRL} These proposed pairing states are robust even in the presence of the imbalanced spin density, i.e., the ``magnetized'' or imbalanced superfluid. Studies on the thermodynamic stability of such exotic pairing states have a long history.~\cite{casalbuoni,matsuda}

Recently, the superfluid phase diagram for the homogeneous system has been extended to the BCS-BEC crossover regime by a number of authors.~\cite{chienPRL06,iskin,mannarelli,sheehy,yang06,hu,paoPRB,son,he1,he2,huang,chenPRA06,hePRA07,chien06,gu,parish} Here, the phase diagram is constructed in a plane of the temperature $T$ and the population imbalance $P$. The PS appears in the BEC side. In the deeper BEC limit, the homogeneous BEC superfluid of the boson-fermion mixture is favored.~\cite{pieri06,pieri07} The FFLO state becomes thermodynamically stable in a narrow window in the BCS side. These studies take into account only one particular form of the FFLO pairing that has a single center-of-mass momentum vector ${\bf Q}$, i.e., $\Delta ({\bf r}) \!=\! \Delta _0 {\rm e}^{i{\bf Q}\cdot{\bf r}}$. However, it should be emphasized that for an arbitrary value of $P$, the oscillation of the stable FFLO phase can be described using the multiple vectors ${\bf Q}$ and $-{\bf Q}$, i.e., $\Delta({\bf r}) \!=\!\Delta _0 \cos{({\bf Q}\cdot{\bf r})}$.~\cite{lo,machida} This effect has not been considered in previous works, except for Ref.~\cite{yoshida} which considers higher harmonics of ${\bf Q}$. The generalized FFLO phase may compete with the PS in the remaining area of the phase diagram.~\cite{yoshida} 

The presence of a trap potential, which is used to capture atomic gas in actual experiments, may lead to a different situation. The simplest way to take account of the trap is to employ a local density approximation (LDA), which is achieved by replacing the chemical potential with the local quantity including the trap potential. The LDA calculation~\cite{yi,yi06,haquea,haquea07,silva,gubbels,martikainen,chienPRA06,chienPRL07} predicts that the PS state, i.e., the BCS state surrounded by the spin-polarized normal domain, is favored under a realistic condition with a large number of particles and a fully three-dimensional trap. However, we should mention that the LDA eliminates the FFLO pairing state, which is one of the possible candidates for the ground state, because of the lack of the gradual spatial variation of the pairing field. Taking into account the gradient effect, full numerical analysis has been performed by several authors in the weak-coupling BCS regime~\cite{MMI1,MMI3,castorina} and at the unitary limit.~\cite{kinnunen,jensen,liu07} They predict the stability of the FFLO oscillation, which cannot be described in terms of a single ${\bf Q}$.

The aim of this paper is twofold. The first goal is to clarify the ground state of trapped fermions with population imbalance. Previously in our series of papers,~\cite{MMI1,MMI3} we presented numerical results  in the weak-coupling BCS regime. The current work covers a wider region, including the BEC side of a resonance in the plane of the population imbalance $P$ and the dimensionless parameter $k_{\rm F}a$, where $k_{\rm F}$ is the Fermi wave number. To address such a problem, we start with the Bogoliubov-de Gennes (BdG) equation,~\cite{degennes} which includes a contribution from the gradient effect of the pairing field and describes the physics in the atomic scale $\sim\! k^{-1}_{\rm F}$. 

The second goal of the present paper is to discuss how the quasi-particle structure in the BCS-BEC crossover regime under population imbalance affects the RF spectroscopy, which has been experimentally performed by Schunck {\it et al}.~\cite{mit4} 

Also, we investigate how a large particle number $N\!=\!\mathcal{O}(10^5)$ changes the FFLO oscillation. In this work, we fully solve the BdG equation coupled with the regularized gap equation and number equation, where the contributions from the higher energy are supplemented by the LDA. This hybrid calculation enables us to demonstrate the stability of the FFLO modulation at the quantitative level, which is comparable to the results of recent experiments.~\cite{rice1,rice2,mit1,mit2,mit3,mit4} The numerical results indicate that FFLO modulation exists even in the vicinity of a Feshbach resonance $1/k_{\rm F}a \!\sim\! -0.5$, which has been already observed experimentally.~\cite{mit1}

This paper is organized as follows. In \S~2, we derive the BdG equation coupled with the regularized gap equation on the basis of the single channel model. Also, we show the numerical results on the $1/k_{\rm F}a$ dependence of basic physical quantities in the BCS-BEC crossover regime without population imbalance. We present the ground-state structures of the imbalanced system in \S~3, where the spatial profiles of the pairing field and densities in the strongly interacting system ($1/k_{\rm F}|a| \!<\! 1$) are displayed. In addition, we shall present a quantum phase diagram in the $1/k_{\rm F}a$-$P$ plane. In \S 4, we present the numerical results on the local density of states (LDOS) and the RF spectroscopy for the imbalanced superfluid in the BCS-BEC crossover regime. The final section is devoted to conclusions and discussion. In addition, we give supplementary information, e.g., the derivation of the thermal Green's function and the gap equation, in Appendices A and B. 

\section{Theoretical Formulation: Single-Channel Model}

\subsection{Bogoliubov-de Gennes equation}

Let us consider a Fermi gas distributed in two hyperfine spin states ($\sigma \!=\! \uparrow , \downarrow$). The Fermi system across a broad Feshbach resonance, which is realized in $^6$Li or $^{40}$K atoms, can be well described by the single-channel Hamiltonian: 
\begin{eqnarray}
{\mathcal{H}} &=& \int d{\bf r} \int d{\bf r}'
 \left[
	\sum _{\sigma} {\psi}^{\dag}_{\sigma}({\bf r}) 
	H^{(0)}_{\sigma} {\psi} _{\sigma}({\bf r}) \delta({\bf r}-{\bf r}') 
	\right. \nn \\
	&& \left. 
	+ U({\bf r}-{\bf r}')
	{\psi}^{\dag}_{\uparrow}({\bf r}) {\psi}^{\dag}_{\downarrow}({\bf r}')
	{\psi}_{\downarrow}({\bf r}') {\psi}_{\uparrow}({\bf r})
\right],
\label{eq:original}
\end{eqnarray}
with the creation and annihilation operators of fermions, ${\psi}^{\dag}_{\sigma}({\bf r}) $ and ${\psi}_{\sigma}({\bf r}) $. The single-particle Hamiltonian is given by
\begin{eqnarray}
H^{(0)}_{\sigma}({\bf r}) =  - \frac{1}{2M} \nabla^2 +V({\bf r}) - \mu _{\sigma},
\label{eq:single}
\end{eqnarray} 
where atoms with mass $M$ are trapped by a harmonic potential $V({\bf r})$. Throughout this paper, we set $\hbar \!=\! k_B \!=\! 1$. The interatomic interaction potential is $U({\bf r}-{\bf r}')$ and the chemical potential of two species is $\mu _{\uparrow, \downarrow} \!=\! \mu \pm \delta\mu$, where, without the loss of generality, we set $\mu _{\uparrow} \!\ge\! \mu _{\downarrow}$, i.e., the spin up (spin down) is the majority (minority) component. 

Following the procedure described in Appendix A, the Bogoliubov-de Gennes equation is given as
\begin{eqnarray}
\left[
\begin{array}{cc}
\mathcal{K}_{\uparrow}({\bf r}) & \Delta({\bf r}) \\
\Delta^{\ast}({\bf r}) & -\mathcal{K}^{\ast}_{\downarrow}({\bf r})
\end{array}
\right]
\left[
\begin{array}{c} u_{\nu}({\bf r}) \\ v_{\nu}({\bf r}) \end{array}
\right] = E_{\nu}
\left[
\begin{array}{c} u_{\nu}({\bf r}) \\ v_{\nu}({\bf r}) \end{array}
\right] ,
\label{eq:bdgeq}
\end{eqnarray}
where the diagonal element is obtained from $\mathcal{K}_{\sigma}({\bf r},{\bf r}') \!\equiv\! \delta ({\bf r}-{\bf r}')\mathcal{K}_{\sigma}({\bf r})$ in eq.~(\ref{eq:matrix2}). This equation describes the quasi-particle state with eigenfunction $[u_{\nu},v_{\nu}]$ and eigenenergy $E_{\nu}$ under the pairing field $\Delta$ and Hartree potential $g\rho _{\sigma}$. Here, the interparticle interaction is characterized by the $s$-wave scattering, $U({\bf r}-{\bf r}') \!=\! g\delta({\bf r}-{\bf r}')$, whose ``bare'' coupling constant is $g \!=\! 4\pi a/M$ with $s$-wave scattering length $a$. Hereafter, the interaction strength of the system is characterized using the dimensionless form $k_{\rm F}a$ with the Fermi wave number $k_{\rm F} \!\equiv\! \sqrt{2ME_{\rm F}}$. $E_{\rm F} $ is the Fermi energy in a noninteracting Fermi gas, whose definition is given in \S~2.2.

The bare coupling constant, however, provides two singular contributions to the BdG equation: (i) the ultraviolet (UV) divergence of the pair potential $\Delta ({\bf r})$ and (ii) the divergence of the Hartree potential at the unitary limit $k_{\rm F}a \!\rightarrow\! \pm\infty$. It is known~\cite{huang87,randeria90,bruun} that the UV divergence can be renormalized by replacing the bare coupling constant $g$ with the effective constant $\tilde{g}({\bf r})$ in the gap equation. The explicit form of the regularized gap equation~\cite{bulgac,grasso} is given as
\begin{eqnarray}
\Delta({\bf r}) = \tilde{g}({\bf r}) \sum _{\nu} u_{\nu}({\bf r}) v^{\ast}_{\nu}({\bf r})f_{\nu},
\label{eq:gapequation}
\end{eqnarray}
where the renormalized coupling constant $\tilde{g}({\bf r})$ is given by 
\begin{eqnarray}
\frac{1}{\tilde{g}({\bf r})} = 
\frac{1}{g}+\frac{Mk_{\rm c}({\bf r})}{2\pi^2} \left[ 1 - \frac{k_{\rm F}({\bf r})}{2k_{\rm c}({\bf r})} 
\ln{\frac{k_{\rm c}({\bf r})+k_{F}({\bf r})}{k_{\rm c}({\bf r})-k_{\rm F}({\bf r})} }\right].
\label{eq:geff1}
\end{eqnarray}
Here, the Fermi distribution function is $f_{\nu} \!\equiv\! f(E_{\nu}) \!=\! 1/ ({\rm e}^{E_{\nu}/T}+1)$. The summation in eq.~(\ref{eq:gapequation}) is carried out for all the eigenstates with both positive and negative eigenenergies, whose details are described in Appendix A. The above gap equation is now free from the energy cutoff $E_{\rm c}$. For $E_{\rm c} \!\gg\! E_{\rm F}$, the expression for $\tilde{g}({\bf r})$ coincides with that obtained from the two-body $T$-matrix in the absence of the medium~\cite{randeria90}. Here, $k_{\rm F}({\bf r})$ and $k_{\rm c}({\bf r})$ are the local wave vectors defined by the local Fermi and cutoff energies, respectively: 
\begin{subequations}
\label{eq:localenergy}
\begin{eqnarray}
E_{\rm F} ({\bf r}) = \mu -  V({\bf r}) , 
\end{eqnarray}
\begin{eqnarray}
E_{\rm c} ({\bf r}) = \frac{k^2_{c}({\bf r})}{2M} + V({\bf r}) - \mu.
\end{eqnarray}
\end{subequations}
Note that even if the gap equation with the bare coupling constant $g$ is singular, the effective constant $\tilde{g}({\bf r})$ yields a nonsingular negative value in an extensive region, ranging from the unitary limit to the deep BEC limit.

For the second divergent behavior (ii), it is known~\cite{baker,heiselberg} that the divergent term at the unitary limit can be renormalized if we consider the many-body contributions beyond the mean-field self-energy, where the system behaves as a Fermi liquid with an effective mass. Hence, we remove the singularity by neglecting the Hartree term, that is, the diagonal elements in the BdG equation (eq.~(\ref{eq:bdgeq})) are replaced by the single-particle Hamiltonian in eq.~(\ref{eq:single}): $\mathcal{K}_{\sigma}({\bf r}) \!=\! H^{(0)}_{\sigma}({\bf r})$. 

In summary, the renormalized coupling constant in eq.~(\ref{eq:geff1}) and the neglect of the Hartree potential lead to the regularization of the BdG formalism. The BdG equation (eq.~(\ref{eq:bdgeq})) is self-consistently coupled with the gap equation (eq.~(\ref{eq:gapequation})) and the number equation
\begin{eqnarray}
N=\sum _{\sigma}N_{\sigma}= \int d{\bf r} \sum _{\sigma}\rho _{\sigma}({\bf r}),
\label{eq:numbereq}
\end{eqnarray}
where 
the particle density in each spin state is obtained from Eqs.~(\ref{eq:hartree}) and (\ref{eq:bogo}) by
\begin{subequations}
\label{eq:defrho}
\begin{eqnarray}
\rho _{\uparrow}({\bf r}) = \sum _{\nu } |u_{\nu}({\bf r})|^2 f_{\nu}, 
\label{eq:rhou} 
\end{eqnarray}
\begin{eqnarray}
\rho _{\downarrow}({\bf r}) = \sum _{\nu} |v_{\nu}({\bf r})|^2 (1- f_{\nu}). 
\label{eq:rhod} 
\end{eqnarray}
\end{subequations}
This formalism is now free from any divergence and provides a qualitative expression for strongly interacting Fermi systems in the BCS-BEC crossover regime.~\cite{leggett} Note that this equation within the single-channel model gives equivalent results to those obtained from another mean-field theory based on the fermion-boson model in the case of a broad resonance.~\cite{ho} Also, in the deep BEC limit ($1/k_{\rm F}a \!\rightarrow \! + \infty$), the BdG equation (eq.~(\ref{eq:bdgeq})) can be mapped to the Gross-Pitaevskii equation with a small parameter $\Delta/|\mu|$,~\cite{pieri} where $\Delta$ describes the wave function of the condensed molecular bosons. As we shall show later, the chemical potential at the unitary limit $1/k_{\rm F}a\!\rightarrow\!0$ is estimated as $\mu/E_{\rm F} \!=\! 1+\beta$ with $\beta\!=\!-0.4$ on the basis of the current theory, which is comparable to the recent experimental result of $\beta \!=\! -0.54$.~\cite{rice1,stewart}

In performing the numerical calculation, we compute the gap equation (eq.~(\ref{eq:gapequation})) by the following hybrid procedure: $\Delta({\bf r})\!=\! \Delta _{\rm BdG}({\bf r}) + \Delta _{\rm LDA}({\bf r})$. The first term is composed of the contributions from low-energy eigenstates $|E_{\nu}| \!<\! E^{({\rm BdG})}_{c}$ obtained from the exact diagonalization of the BdG equation. The quantity $\Delta _{\rm LDA}$ with the higher-energy contribution above $E^{({\rm BdG})}_{\rm c} \!<\! E_{\nu} \!<\! E_{\rm c}$ is supplemented by the LDA, whose explicit expression is given by
\begin{eqnarray}
\Delta _{\rm LDA} ({\bf r}) = \tilde{g}({\bf r}) \int^{\infty}_{p^{({\rm BdG})}_{\rm c}} \frac{d{\bf p}}{(2\pi)^3} 
\frac{\Delta({\bf r})}{2E({\bf p},{\bf r})} \hspace{10mm} \nn \\ 
\times [ f(E_{\uparrow}({\bf p},{\bf r})) + f(E_{\downarrow}({\bf p},{\bf r})) - 1 ],
\label{eq:gaplda}
\end{eqnarray}
with $p^{({\rm BdG})}_{\rm c} \!\equiv\! \sqrt{2M E^{({\rm BdG})}_{\rm c}}$. Here, we set $E_{\uparrow,\downarrow}({\bf p},{\bf r})\!=\!E({\bf p},{\bf r})\mp \delta\mu$ with $E({\bf p},{\bf r})\!=\! \sqrt{[\epsilon({\bf p},{\bf r})]^2+|\Delta({\bf r})|^2}$, $\epsilon({\bf p},{\bf r})\!=\!p^2/2m - \mu$, and $E_{\rm c}\!=\!p^{2}_{\rm c}/2m - \mu$. Also, the high energy contribution to each spin density is expressed within the LDA as $\rho \!=\! \rho ^{({\rm BdG})}_{\sigma} + \rho ^{({\rm LDA})}_{\sigma}$ with
\begin{eqnarray}
\rho ^{({\rm LDA})}_{\uparrow,\downarrow}({\bf r}) = \frac{1}{2}\int^{\infty}_{p^{({\rm BdG})}_{\rm c}} \frac{d{\bf p}}{(2\pi)^3} 
\left[ 
\left\{ 1 + \frac{\epsilon({\bf p},{\bf r})}{E({\bf p},{\bf r})} \right\} \right. \hspace{7mm} \nn \\
\left.
\times f(E_{\uparrow,\downarrow}({\bf p},{\bf r})) 
+ \left\{ 1 - \frac{\epsilon({\bf p},{\bf r})}{E({\bf p},{\bf r})} \right\} f(-E_{\downarrow,\uparrow}({\bf p},{\bf r})) 
\right]. 
\label{eq:rholda}
\end{eqnarray}
Note that the spatial variation of the pairing field is mainly determined by the contributions from the eigenstates with energy close to the Fermi energy, while the eigenstates with the higher energy may be described within the semiclassical approximation, i.e., the LDA. This hybrid procedure has also been used in the numerical analysis of the thermodynamic quantities at the unitary limit.~\cite{liu07}

\subsection{Calculated system}

We numerically solve the BdG equation (eq.~(\ref{eq:bdgeq})), which is self-consistently coupled with the gap equation (eq.~(\ref{eq:gapequation})). The theory takes account of the trap potential and the mismatch of Fermi surfaces $\delta\mu \!\equiv\! (\mu _{\uparrow}-\mu_{\downarrow})/2$. At each iteration step, the chemical potential $\mu$ is adjusted to fix the total particle number defined in eq.~(\ref{eq:numbereq}). In the current work, we consider a cylindrical symmetric system with trap potential $V({\bf r}) \!=\! \frac{1}{2}M\omega^2 r^2$ ($r^2 \!=\! x^2+y^2$), and impose a periodic boundary condition with periodicity $Z \!=\! 3d$ ($d^{-1} \!\equiv\! \sqrt{M\omega}$) along the $z$-direction. Under such cylindrical symmetry, the quasi-particle wave functions are written as $u_{\nu} ({\bf r}) \!=\! u_{\nu}(r) {\rm e}^{i(q_{\theta} \theta + q_z z)}$ and $v_{\nu} ({\bf r}) \!=\! v_{\nu}(r) {\rm e}^{i(q_{\theta} \theta + q_z z)}$ with quantum numbers along the azimuthal- and $z$-axis: $q_{\theta} \!=\! 0, \pm 1, \pm 2, \cdots$ and $q_z \!=\! 0, \pm 2\pi/Z, \pm 4\pi /Z , \cdots$.

The BdG matrix in eq.~(\ref{eq:bdgeq}) is then transformed by spatial discretization into a banded matrix with respect to the radial axis, which can be solved using the Lanczos/Arnoldi algorithm implemented in the ARPACK libraries.~\cite{arpack} Throughout this paper, we use total particle numbers of $N \!=\! 3,000$ and $150,000$. The corresponding Fermi energies in $\Delta \!=\! 0$ are given as $E_{\rm F} \!=\! 32 \omega$ and $154 \omega$, respectively, using the definition $E_{\rm F}/\omega \!=\! (30\pi n_z /16)^{2/5}$ with $n_z \!\equiv\! N/Z$. Throughout this paper, we set $E^{({\rm BdG})}_{\rm c}\!=\!150 \omega \!=\! 4.7 E_{\rm F}$ for the case of $N \!=\! 3,000$ atoms and $E^{({\rm BdG})}_{\rm c}\!=\!200 \omega \!=\! 1.3 E_{\rm F}$ for $N \!=\! 150,000$ to limit computation time. However, the higher-energy contributions up to $E_{\rm c} \!=\! 1000\omega$ are supplemented within the LDA, as shown in Eqs.~(\ref{eq:gaplda}) and (\ref{eq:rholda}).

\subsection{Calculations for balanced systems}

\begin{figure}[b!]
\begin{center}
\includegraphics[width=0.8\linewidth]{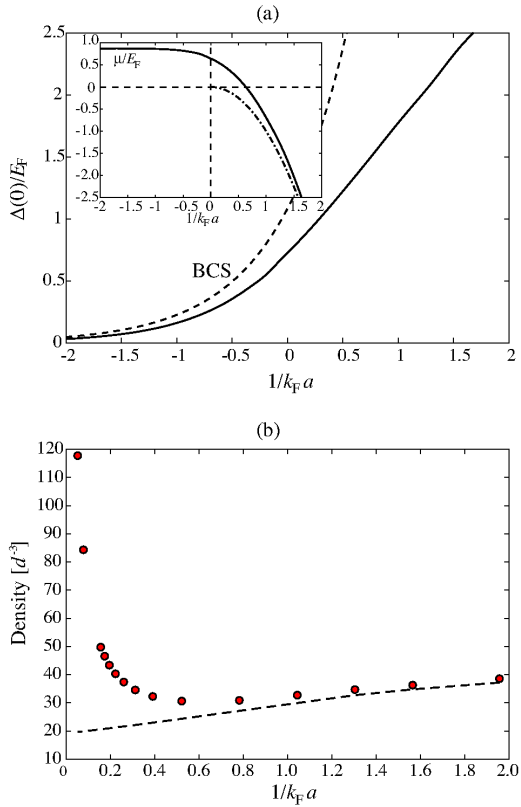}
\end{center}
\caption{(a) Maximum values of pairing amplitude in balanced population ($P\!=\!0$) at $T\!=\!0$ and $N\!=\! 3,000$ as a function of $1/k_{\rm F}a$. The dashed line is the BCS form described in the text. The inset displays the chemical potential shift. The dash-dotted line corresponds to one-half of the binding energy $E_b/2E_{\rm F}$. (b) The order parameter of the molecular bosons $2|\Psi _{\rm BEC}(r\!=\!0)|^2$ (circles) is compared with the total density $\rho (r\!=\!0)$ (dashed line). The definition for $\Psi _{\rm BEC}$ is given in the text.  
}
\label{fig:gap_balance}
\end{figure}

Before discussing the numerical results for superfluid states with population imbalance, let us present the basic properties of balanced superfluids, described within the above mentioned mean-field theory. First, in Fig.~\ref{fig:gap_balance}(a), we show the pairing amplitude $\Delta _0/E_{\rm F}$ as a function of $1/k_{\rm F}a$, where $\Delta _0$ is the maximum value of the pairing field at zero temperatures. It is found that in the weak coupling limit $1/k_{\rm F}a \!<\! -1$, the pairing field can be asymptotically described by the standard BCS relation, $\Delta _0 /E_{\rm F} \!=\! 8{\rm e}^{-2-\pi/k_{\rm F}|a|}$, while the pairing in the opposite limit becomes a wave function of tightly bound molecular bosons, i.e., the order parameter of BEC. In this BEC limit, it is known that the BdG equation in the single-channel model can be mapped into the Gross-Pitaevskii equation for molecular bosons~\cite{pieri} where the fermionic chemical potential becomes one-half of the binding energy of the pairs $E_b/E_{\rm F} \!=\! - 2/(k_{\rm F}a)^2$, shown in the inset of Fig.~\ref{fig:gap_balance}(a). The order parameter of the molecular bosons is expressed as $\Psi _{\rm BEC} ({\bf r}) \!=\! \sqrt{\frac{M^2a}{8\pi}}\Delta ({\bf r})$, corresponding to the total fermionic density $2|\Psi _{\rm BEC}({\bf r})|^2 \!=\! \rho ({\bf r})$. As shown in Fig.~\ref{fig:gap_balance}(b), this asymptotic behavior can be confirmed by the direct calculation using the BdG equation. 

The intermediate region nearby $1/k_{\rm F}a\!=\!0$ is smoothly connected from the BCS to BEC limits. Then, the pairing field $\Delta$ obtained from the single-channel model describes the order parameter composed of the fermionic pairs and the wave function of the molecular BEC. The pairing amplitude at the unitary limit is $\Delta _0/E_{\rm F} \!=\! 0.7$, which is overestimated in comparison with $\Delta _0/E_{\rm F} \!=\! 0.5$ in the strong-coupling theory.~\cite{gorkov,bulgacPRL06} It is also found that the coherence length is saturated toward the length scale of the interatomic spacing $\xi _0 k_{\rm F} \!=\! 2E_{\rm F}/\Delta _0 \!=\! \mathcal{O}(1)$ as $k_{\rm F}a$ approaches the BEC limit $1/k_{\rm F}a \!\rightarrow\! \infty$.

\section{Ground States in the Imbalanced System at $T\!=\! 0$}

\subsection{Superfluid states in weak-coupling BCS regime}


\begin{figure}[h!]
\begin{center}
\includegraphics[width=0.85\linewidth]{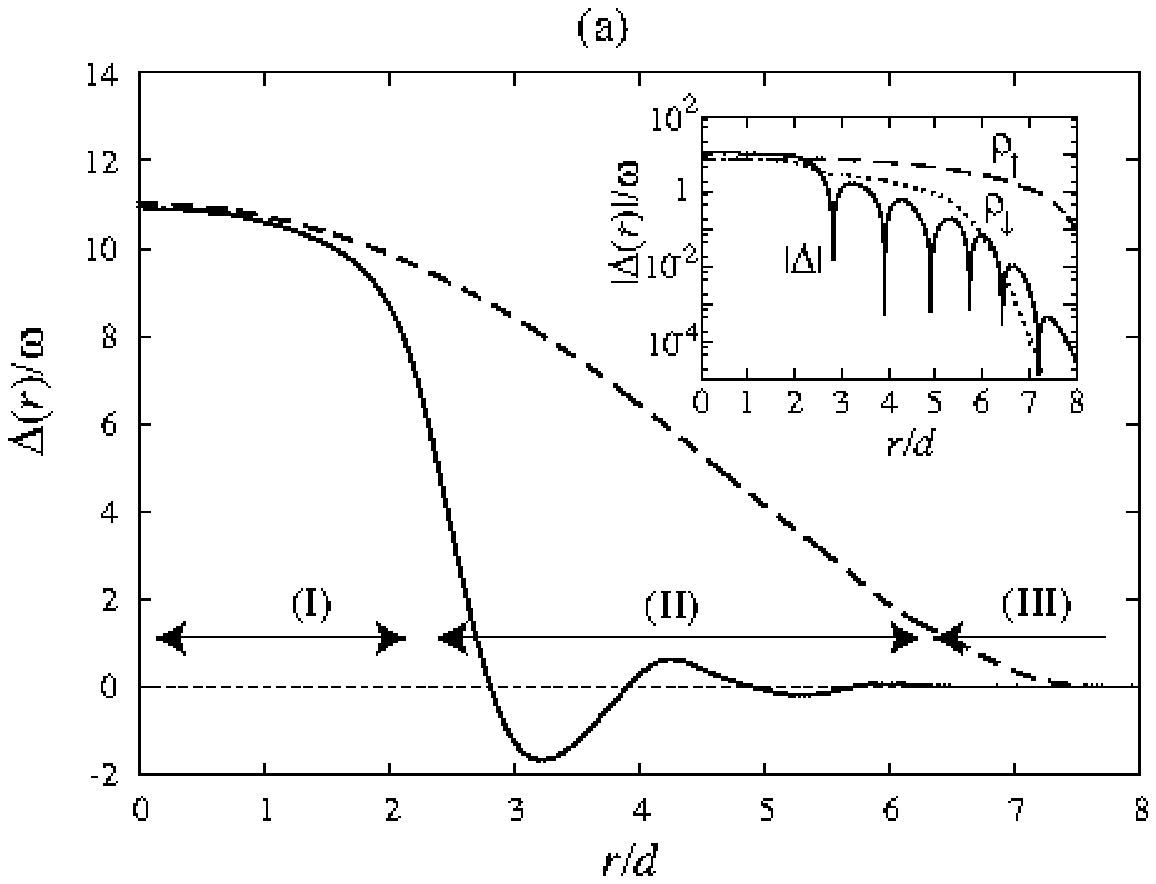} 
\includegraphics[width=0.85\linewidth]{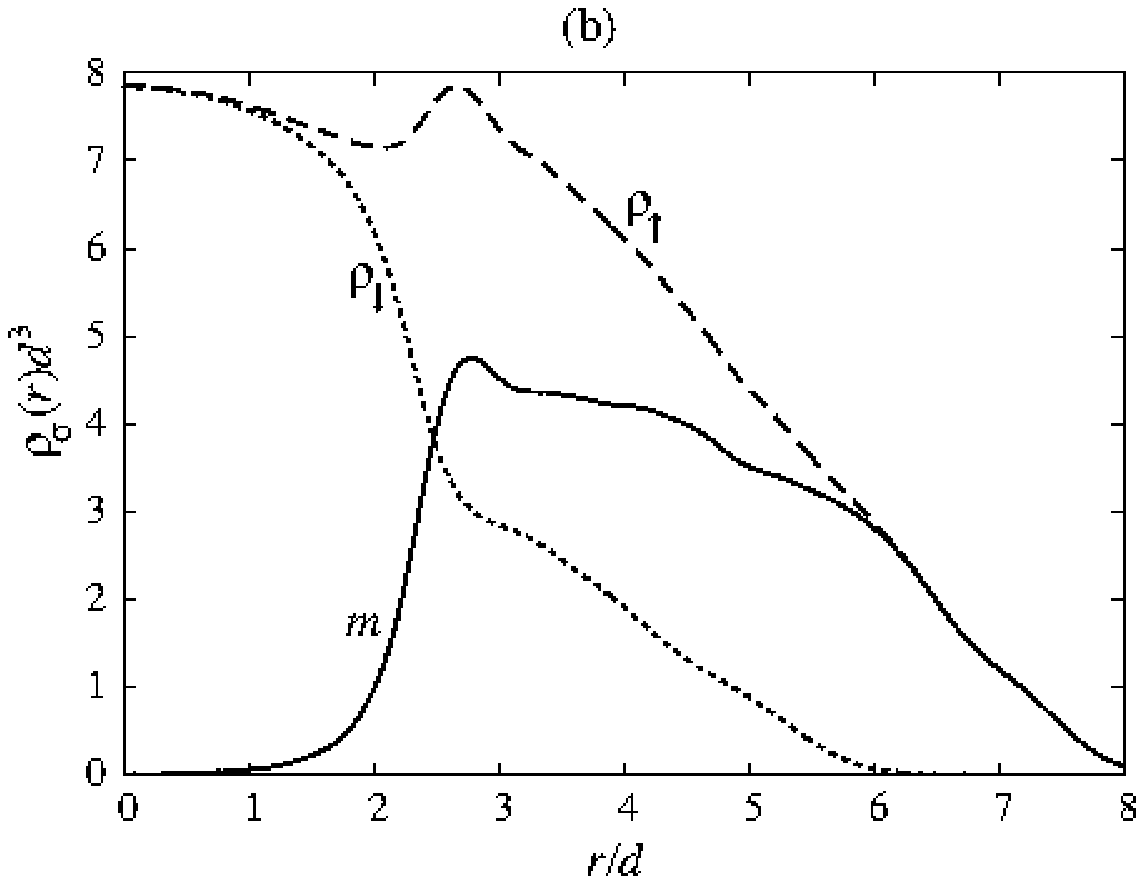} 
\includegraphics[width=0.85\linewidth]{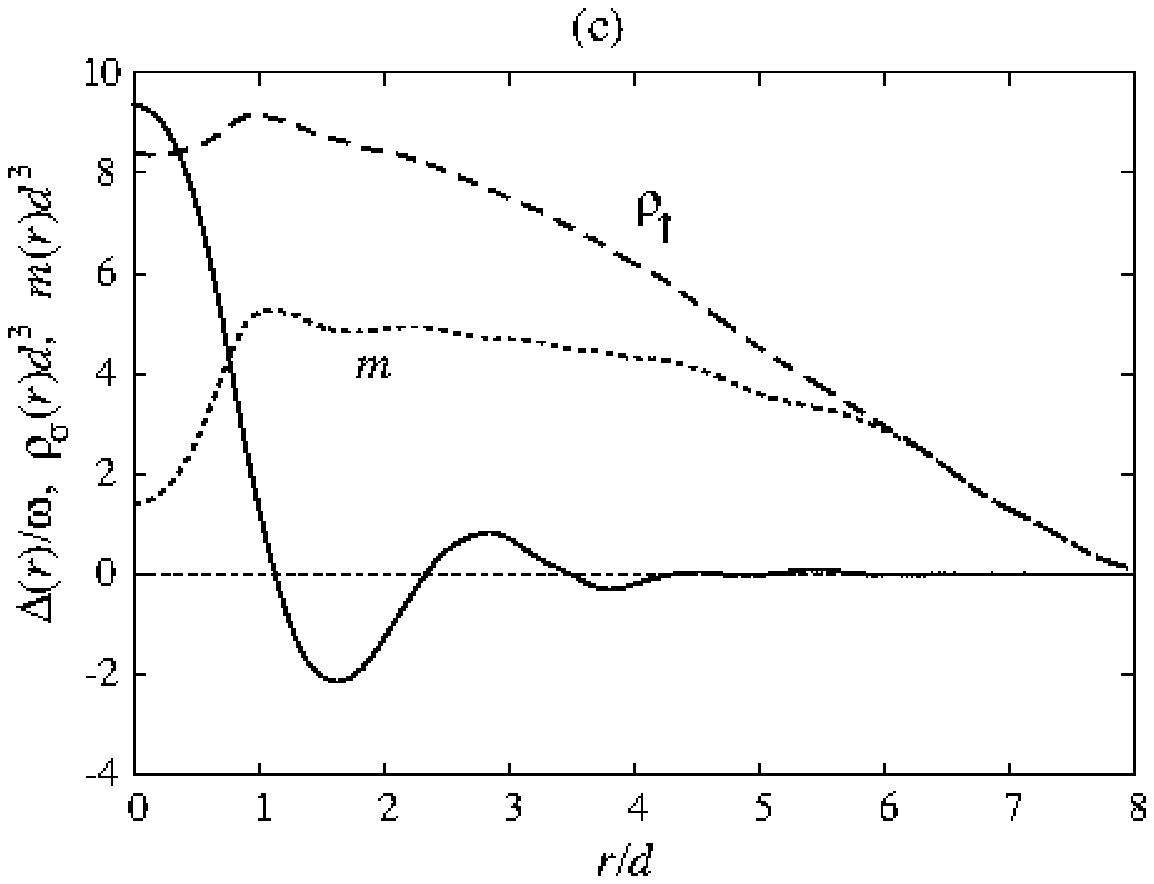} 
\end{center}
\caption{Spatial profiles of (a) the pairing field at $P\!=\!0$ (dashed line) and $P\!=\!0.48$ (solid line) and (b) the corresponding density profiles, where the dashed, dotted, and solid lines denote the majority and minority densities and the local magnetization, respectively. The inset in (a) shows $\Delta(r)$ (solid line) and $\rho _{\sigma}(r)$ (dashed and dotted lines) with a logarithmic scale. In (c), $\Delta (r)$, $m(r)$, and $\rho _{\uparrow}(r)$ at $P\!=\!0.58$ are displayed with solid, dotted, and dashed lines. All results are at $T\!=\!0$ and $1/k_{\rm F}a \!=\! -0.52$. 
}
\label{fig:gap_g-3T0}
\end{figure}

We now consider the imbalanced case for $N \!=\! 3,000$ fermions trapped by a cylindrical potential at $T\!=\!0$. Figure \ref{fig:gap_g-3T0}(a) shows the spatial profiles of the pairing field at $P\!=\!0$ (dashed line) and $P\!=\!0.48$ (solid line) in the weak-coupling BCS side of a resonance for $1/k_{\rm F}a \!=\! -0.52$. For this coupling constant, it is found that the quantum phase transition from the superfluid state to the normal state is induced at the critical population imbalance $P_{\rm c} \!=\! 0.61$. It can be seen from Fig.~\ref{fig:gap_g-3T0}(a) that in the central region labeled (I), the population imbalance does not affect the pairing field. In contrast, the superfluid pairing field is quenched in the outside region labeled (III). The pairing field in the intermediate region (II) yields the spatial oscillation, i.e., the FFLO modulation, where the amplitude gradually decreases toward the edge of the cloud.

These characteristics of $\Delta (r)$ are reflected by the density profiles of each spin component displayed in Fig.~\ref{fig:gap_g-3T0}(b), where we define the local population difference, called the local ``magnetization'', as
\begin{eqnarray}
m({\bf r}) \equiv \rho _{\uparrow}({\bf r})-\rho _{\downarrow}({\bf r}).
\end{eqnarray}
Spin states in region (I) attract each other, and the magnetization is excluded in order to obtain the full condensation energy. In the intermediate region (II) the gap function changes its sign, allowing it to accommodate the excess majority species. This is indeed a characteristic of the FFLO state; The accumulation of excess majority species at $T\!=\!0$ results from the topological structure of $\Delta(r)$. The quasi-particles across the FFLO node undergo a $\pi$-phase shift of the pair potential, allowing them to form a mid gap state that is spatially bound there, called the Andreev bound state in more general contexts.~\cite{machida,kashiwaya,MMI2} The energy of this state is situated in the middle of the energy gap, and the mismatch of the Fermi surface causes the difference in the occupation of this bound state, leading to the local magnetization around the nodes, as seen in Fig.~\ref{fig:gap_g-3T0}(b). Here, a sufficient amount of the minority component remains for pairing with the majority component, i.e., the local magnetization is not fully polarized. These features are revealed by the bimodal distribution of the minority component (see around $r/d\sim 3$ in Fig.~\ref{fig:gap_g-3T0}(b)). 

\begin{figure}[t!]
\begin{center}
\includegraphics[width=0.9\linewidth]{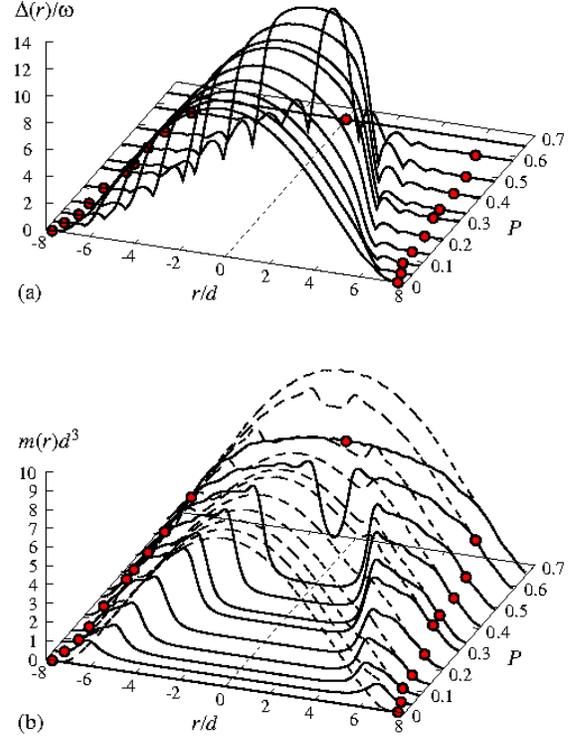} 
\end{center}
\caption{
(a) Spatial distributions of $|\Delta (r)|$ at $1/k_{\rm F}a \!=\! -0.52$ and $T\!=\! 0$ for various values of $P$. (b) Corresponding local magnetization $m(r)$ (solid lines) and the majority density $\rho _{\uparrow}(r) $ (dashed lines). The circles correspond to the radii of the pairing field, $R_{\rm c}$ defined as $\Delta(R_{\rm c})/\Delta _0 \!=\! 10^{-3}$.
}
\label{fig:3dgap_g-3T0}
\end{figure}

Figure \ref{fig:gap_g-3T0}(c) shows the gap and density profiles in the vicinity of the critical population imbalance $P/P_{\rm c} \!=\! 0.95$. Here, the local magnetic moment $m(r)$ governs the entire region of the system, where the ``empty core'' in the central region of the local magnetization vanishes. This leads to the quenching of the balanced BCS pairing even at $r\!\sim\! 0$. As seen in Fig.~\ref{fig:gap_g-3T0}(c), however, the superfluidity remains robust up to the edge of the minority component $r \!\sim\! 6.5d$ by forming the FFLO pairing, accompanied with partially polarized spins.

In Fig.~\ref{fig:3dgap_g-3T0} we display the $P$-dependence of the pairing field and local magnetization at $T\!=\!0$ and $1/k_{\rm F}a \!=\! -0.52$. With increasing $P$, the oscillating region becomes wider towards the central region, while the radii of the pairing field $R_{\rm c}$ keep a constant value $R_{\rm c}/d \!\sim\! 7$ up to the critical population imbalance of $P_{\rm c} \!=\! 0.61$. This FFLO pairing state is not describable with the LDA where the superfluidity is localized in the central region (I) and the regions (II) and (III) are regarded as being in the normal state, i.e., the BCS-normal PS state.~\cite{yi,yi06,haquea,silva} It is found that this oscillating pairing state becomes robust up to the critical population imbalance $P_{\rm c} \!=\! 0.6$ at $T\!=\!0$ and $1/k_{\rm F}a\!=\!-0.52$; The equal population $P\!=\!0$ is the only stable situation for the nonoscillating BCS state. Beyond $P_{\rm c}$, the superfluid state becomes the normal state through a second-order phase transition. 

In the outside region of $R_{\rm c}$, where the gap almost vanishes, the complete spin-polarized state is attained. It should be emphasized again that the central region $r\!\sim\!0$, in which the magnetization is completely excluded, catches a clear signature of the ``balanced'' superfluidity, while the surrounding area with the partially polarized spins also keeps its superfluidity composed of the ``imbalanced'' FFLO pairing.

\subsection{Strong-coupling region: from unitary limit to BEC regime}

\begin{figure}[b!]
\begin{center}
\includegraphics[width=0.9\linewidth]{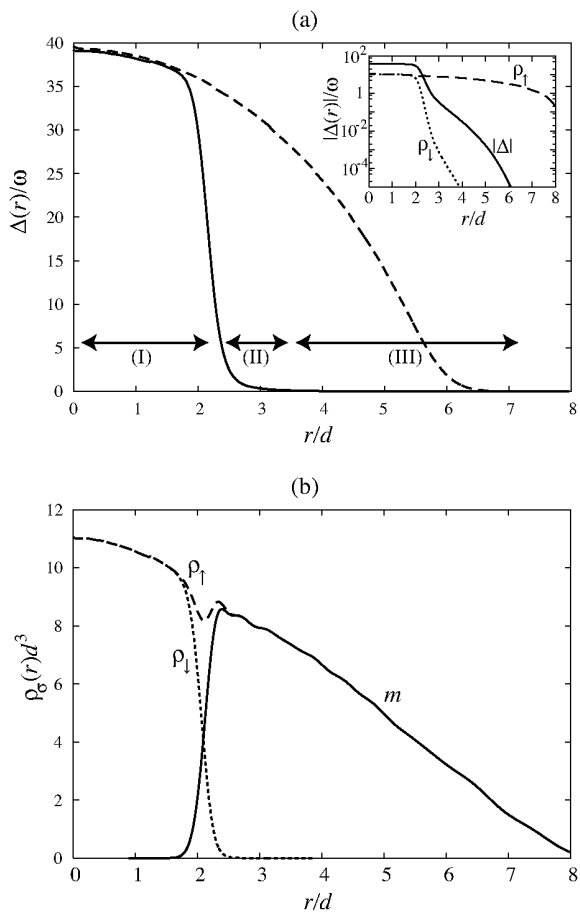} 
\end{center}
\caption{Spatial profiles of (a) the pairing field at $P\!=\!0$ (dashed line) and $P\!=\!0.73$ (solid line) and (b) the corresponding density profiles, where the dashed, dotted, and solid lines denote the majority and minority densities and the local magnetization, respectively. The inset in (a) shows $\Delta(r)$ (solid line) and $\rho _{\sigma}(r)$ (dashed and dotted lines) with a logarithmic scale. All results are for the BEC side $1/k_{\rm F}a \!=\! +0.52$ at $T\!=\!0$.
}
\label{fig:gap_g+3T0}
\end{figure}

Let us now consider the strong-coupling BEC region. The pairing field in the BEC side of the resonance is completely different from that in the BCS side. Figure \ref{fig:gap_g+3T0} shows the spatial profiles of $\Delta(r)$, $\rho _{\sigma}(r)$, and $m(r)$ at $1/k_{\rm F}a \!=\! 0.52$, $P\!=\!0.73$, and $T\!=\! 0$. The superfluid pairing state is still robust in the central region of the system, while the outside region turns to the normal state. The intermediate region (II) smoothly connects the superfluid core (I) and the normal state (III) without any spatial oscillation of the pairing. In contrast with the BCS side shown in Fig.~\ref{fig:gap_g-3T0}, it is seen from the inset of Fig.~\ref{fig:gap_g+3T0}(a) that the intensity of the pairing field exponentially decays toward the edge of the minority component. As seen in Fig.~\ref{fig:gap_g+3T0}(b), the magnetization is completely excluded from the central region (I), while the spins in the outer region (III) are fully polarized. In the strong-coupling regime, the intermediate region (II) plays the role of the domain wall between the superfluid core (I) and the fully polarized normal state (III). This phase-separated profile is commonly seen in other theoretical calculations based on the LDA,~\cite{yi,yi06,haquea,haquea07,silva,gubbels,chienPRA06,chienPRL07} except for the proximity effect of the pairing field in the vicinity of the domain wall.

\begin{figure}[b!]
\begin{center}
\includegraphics[width=0.9\linewidth]{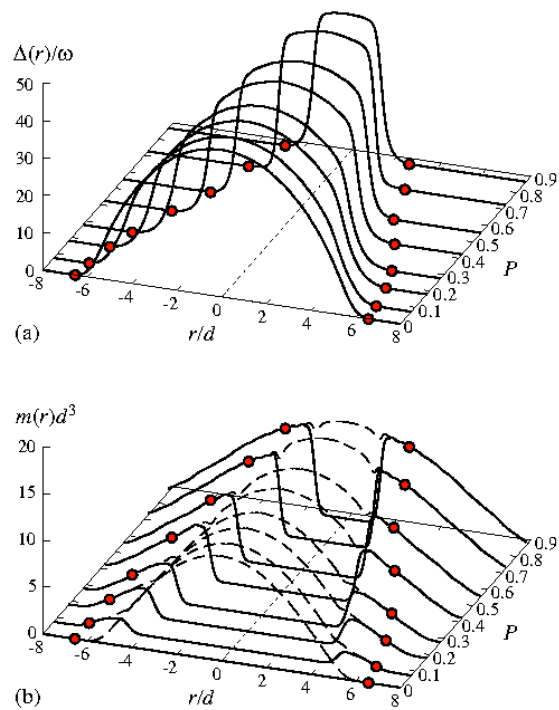} 
\end{center}
\caption{
(a) Spatial distributions of $|\Delta (r)|$ at the BEC side $1/k_{\rm F}a \!=\! 0.52$ and $T\!=\! 0$ for various values of $P$. (b) Corresponding local magnetization $m(r)$ (solid lines) and the majority density $\rho _{\uparrow}(r) $ (dashed lines). The circles correspond to the edge of the pairing field $R_{\rm c}$ defined as $\Delta(R_{\rm c})/\Delta _0 \!=\! 10^{-3}$. 
}
\label{fig:3dgapT0bec}
\end{figure}

Figure~\ref{fig:3dgapT0bec} displays the spatial distributions of the pairing field and the local magnetization as a function of $P$, where the circles denote the edge of the pairing field $R_{\rm c}$ defined as $\Delta (R_{\rm c})/\Delta _0 \!=\! 10^{-3}$. In the BEC side, the critical population imbalance is uniquely determined as $P_{\rm c} \!=\!1$ at zero temperatures. The robustness of the superfluidity is supported by the fact that it is realized by the formation of local pairs, i.e., molecules. 

Also, for the $P$-dependence of the ground state shown in Fig.~\ref{fig:3dgapT0bec}, two clear differences exist between the BEC and BCS regimes; First, in the case of $1/k_{\rm F}a \!=\! 0.52$, the radii of the condensation area $R_{\rm c}$ gradually shrinks as $P$ approaches $P_{\rm c} \!=\! 1$, while $R_{\rm c}$ in the BCS regime has an almost fixed value up to $P_{\rm c}$. Second, a domain composed of ``fully'' polarized spins grows around the empty core for all values of $P$ in the BEC side. In contrast, as seen in Fig.~\ref{fig:3dgap_g-3T0}, the outside area in the BCS side has a sufficiently large region (II), which is composed of ``partially'' polarized spins and the minority spins to take part in the superfluid pairing with the majority spins.

\begin{figure}[b!]
\begin{center}
\includegraphics[width=0.9\linewidth]{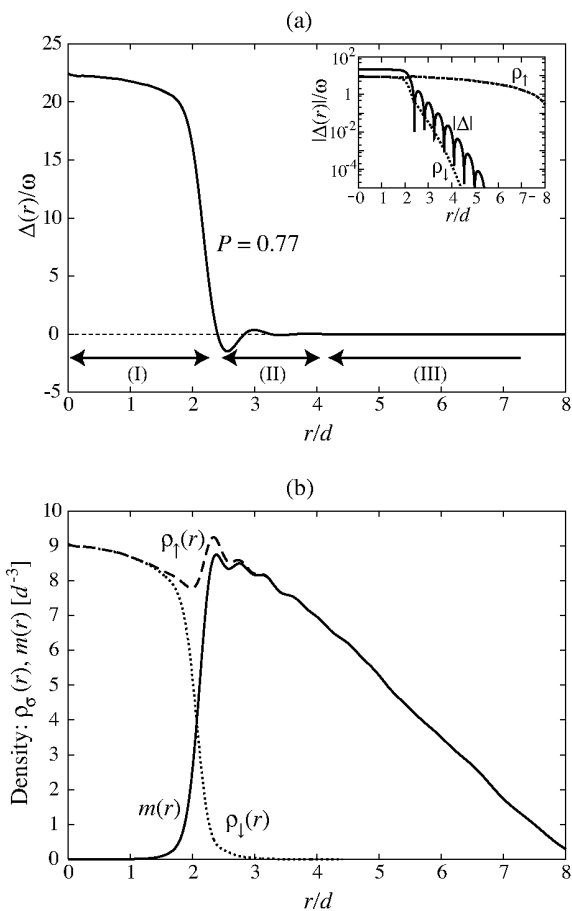} 
\end{center}
\caption{Spatial profiles of (a) the pairing field and (b) the corresponding density profiles at $P\!=\!0.77$ at the unitary limit $1/k_{\rm F}a \!=\! 0$. In (b), the dashed, dotted, and solid lines denote the majority and minority densities and the local magnetization, respectively. The inset in (a) shows $\Delta(r)$ (solid line) and $\rho _{\sigma}(r)$ (dashed and dotted lines) with a logarithmic scale. All results are at $T\!=\!0$.
}
\label{fig:gapT0uni}
\end{figure}

The pairing field and density profiles at the unitary limit are displayed in Fig.~\ref{fig:gapT0uni} where the FFLO oscillation remains in the outside area of the ``core'', which has the balanced spin density. This modulation survives as a proximity effect between the balanced superfluid and the fully polarized domains, analogous to superconductor/ferromagnet interfaces.~\cite{buzdin} Note that the periodicity of the FFLO oscillation may be scaled with the coherence length $\xi _0$. As shown in the inset of Fig.~\ref{fig:gapT0uni}, the FFLO oscillation is completely periodic with periodicity $L \!\sim\! d$, which is comparable to $\xi _0 \!=\! 2.5k^{-1}_{\rm F} \!=\! 0.31d$, i.e., $L \!\sim\! 3\xi _0$. In the BCS side for $k_{\rm F}a \!=\! -0.52$, the FFLO modulation region (II), which becomes wider toward the edge of the cloud, has longer periodicity, $L \!\sim\! 2d \!=\!2.7\xi _0$, e.g., see Fig.~\ref{fig:gap_g-3T0}, where $\xi _0 \!=\! 6k^{-1}_{\rm F} \!=\! 0.75d$. In contrast, the coherence length saturates at the interparticle spacing, $\xi _0 \!\sim\! k^{-1}_{\rm F}$, in the BEC side, in which the oscillation suddenly vanishes. The resulting density profile shown in Fig.~\ref{fig:gapT0uni}(b) is almost unchanged compared with that in the BEC side (See Fig.~\ref{fig:gap_g+3T0}), indicating a phase separation. This is different from the density profile in the BCS side as shown in Fig.~\ref{fig:gap_g-3T0}(b). 

\subsection{Large-$N$ system}

\begin{figure}[b!]
\begin{center}
\includegraphics[width=0.9\linewidth]{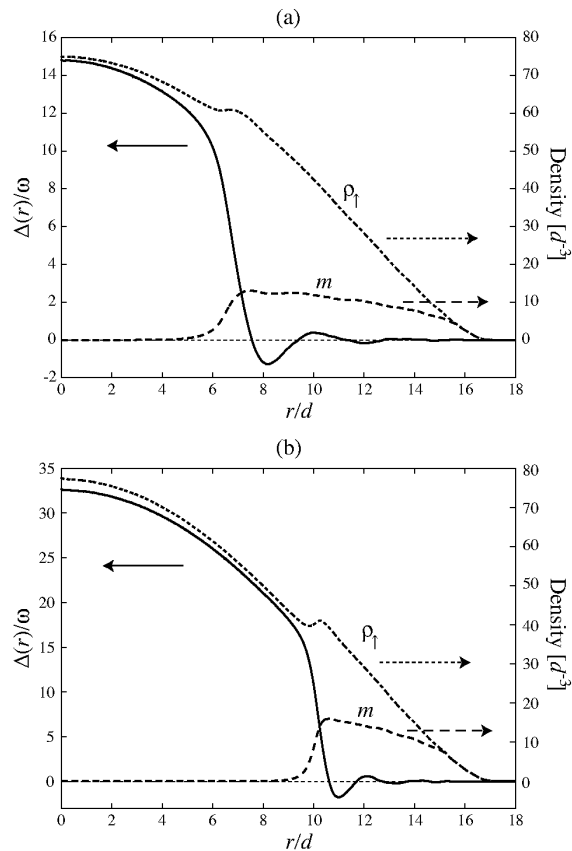} 
\end{center}
\caption{
Spatial profiles of the pairing field (solid line), the majority spin density (dotted line), and the local magnetization (dashed line) in the system with $N \!=\! 150,000$ at (a) $1/k_{\rm F}a \!=\! -1.2$ and $P \!=\! 0.13$ and (b) $1/k_{\rm F}a \!=\! -0.7$ and $P \!=\! 0.12$. All results are at $T\!=\! 0$.
}
\label{fig:largeNg-0.6}
\end{figure}

\begin{figure}[t!]
\begin{center}
\includegraphics[width=0.9\linewidth]{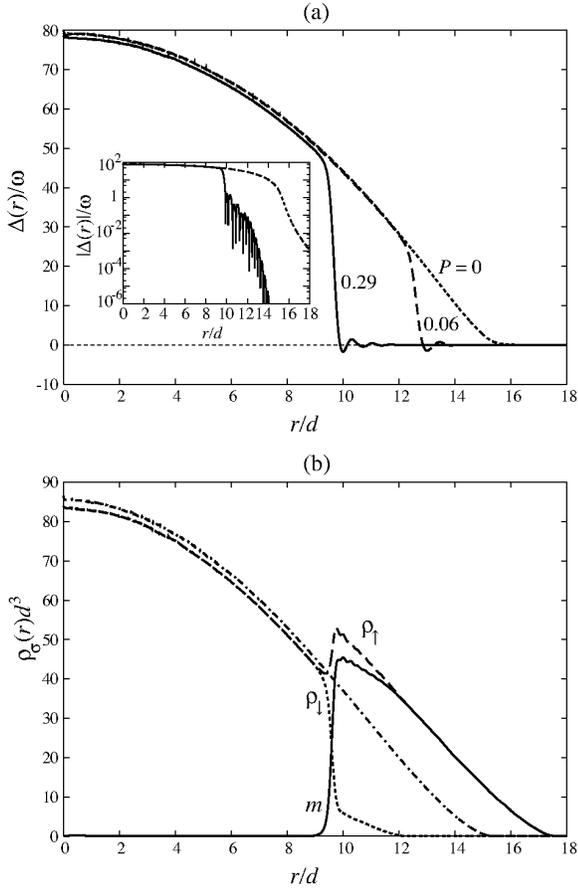} 
\end{center}
\caption{
Spatial profiles of (a) the pairing field and (b) the density in the system with $N \!=\! 150,000$ at $1/k_{\rm F}a \!=\! -0.14$ and $T\!=\! 0$. The dotted, dashed, and solid lines in (a) show the pairing fields at $P \!=\! 0$, $0.06$, and $0.29$, respectively, and those in (b) denote the minority and majority densities, and the local magnetization, respectively. The dot-dashed line in (b) corresponds to the density of the up spins at $P\!=\! 0$. In the inset of (a), $|\Delta(r)|$ at $P\!=\! 0$ (dotted line) and $0.29$ (solid line) are displayed with a logarithmic scale.
}
\label{fig:largeNg-5}
\end{figure}

We now turn to the situation for the realistic particle number $N\!=\!150,000$. The spatial profiles of the pairing field and densities at $1/k_{\rm F}a \!=\! -1.2$ and $-0.7$, and in the vicinity of the unitary limit $1/k_{\rm F}a \!=\! -0.14$ are displayed in Figs.~\ref{fig:largeNg-0.6} and \ref{fig:largeNg-5}, respectively. The tendencies due to the FFLO pairing, which have been already seen in the system with $N\!=\! 3,000$ atoms, are commonly reproduced even in the large-$N$ system. For instance, the outside region $r\!\ge\! 8d$ of Fig.~\ref{fig:largeNg-0.6} shows that the pairing field yields the FFLO oscillation, whose indirect signature at the macroscopic level is the partially polarized spin density. Note that recent experiments~\cite{mit1} have been performed over the wide $k_{\rm F}a$ range, $1/k_{\rm F}|a| \!\le\! 0.5$, in which the FFLO modulation survives. The periodicity $L$ of the oscillation in Fig.~\ref{fig:largeNg-5}(a) has a larger length scale than the interparticle spacing, $L \!\sim\! 4d \!=\! 70 k^{-1}_{\rm F}$, which is comparable to the coherence length $\xi _0$: $L \!=\! 3.3 \xi _0$ with $\xi _0\!=\! 21 k^{-1}_{\rm F}$ and $k^{-1}_{\rm F} \!=\! 0.06d$.

The spatial profiles in the strong-coupling regime are shown in Fig.~\ref{fig:largeNg-5}. The local magnetization is in good agreement with the results of an experiment in which the magnetism was observed by phase-contrast imaging and 3D image reconstruction.~\cite{mit3} While the local magnetization yields a PS-like profile, FFLO modulation occurs in the pairing field. In the strong-coupling regime $1/k_{\rm F}a \!=\! -0.14$, the periodicity of the FFLO oscillation becomes shorter, which is comparable to the interparticle spacing $L \!\sim\! 0.6d \!\sim\! 10.5k^{-1}_{\rm F}$, and its intensity exponentially decays as seen in the inset of Fig.~\ref{fig:largeNg-5}(a). With $\xi _0 \!=\! 3.7k^{-1}_{\rm F}$, the oscillation period is scaled as $L \!\sim\! 2.8\xi _0$. 

In summary, throughout the extensive range of the interaction $k_{\rm F}a \!<\! 0$, the oscillation periodicity $L$ is well scaled with the coherence length $\xi _0$ as $L \!=\! \alpha \xi _0$. The coefficient $\alpha$ is around $3$. We also find that this result is insensitive to the total particle number, i.e., the Fermi wavelength $k^{-1}_{\rm F}$. Surprisingly, it is found that the period is almost unchanged with an increase in the population imbalance. One example of this is displayed in Fig.~\ref{fig:3dgap_g-3T0}(a), where internode spacing is almost fixed for the entire range of $P$. Note that since the FFLO oscillation periodicity in the absence of the trap potential is sensitive to the population imbalance or alternatively to the mismatch of the Fermi surface,~\cite{machida,MMI1} the constancy of $L$ may be peculiar to the finite trap system.  

\subsection{Quantum phase diagram}

\begin{figure}[b!]
\begin{center}
\includegraphics[width=0.9\linewidth]{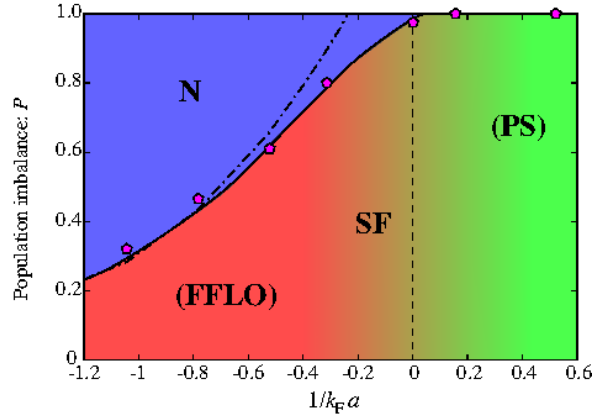}
\end{center}
\caption{
(Color online) Quantum phase diagram in $1/k_{\rm F}a$-$P$ plane. The points denote the estimated phase boundary between the SF and normal (N) phases. The pairing field in the superfluid phase exhibits the spatial oscillation (nonoscillation) for negative (positive) $1/k_{\rm F}a$. The dashed-dotted line is the extrapolation from the results of the weak-coupling limit, $P_{\rm c} \!=\! 1.9 \Delta _0/E_{\rm F}$,~\cite{MMI3} where $\Delta _0$ is the maximum intensity of $\Delta (r)$ at $T\!=\! 0$ and $P\!=\! 0$.
}
\label{fig:phaseT0}
\end{figure}

Let us now summarize the ground state of the imbalanced Fermi system in the BCS-BEC crossover regime by constructing the phase diagram at zero temperatures. Figure \ref{fig:phaseT0} shows the phase diagram in the plane of the dimensionless coupling constant $k_{\rm F}a$ versus the population imbalance $P$ for the system with $N \!=\! 3,000$ atoms. It is important to mention that the critical population imbalance in the weak-coupling limit $1/k_{\rm F}a\!<\!-1$ exponentially depends on $1/k_{\rm F}a$, i.e., $P_{\rm c} \!\propto\! {\rm e}^{-\pi/2k_{\rm F}|a|}$, which results in the linear relationship with the intensity of the pairing field, i.e., $P_{\rm c} \!\propto\! {\Delta_0 \over E_{\rm F}}$. In a previous work,~\cite{MMI3} we found that $P_{\rm c} \!=\! 1.9 \frac{\Delta _0}{E_{\rm F}}$. In Fig.~\ref{fig:phaseT0}, the corresponding exponential line is depicted with $1.9\Delta _0/E_{\rm F}$ where $\Delta _0$ is the maximum gap $\Delta (r \!=\! 0)$ at $T\!=\! 0$ and $P\!=\! 0$. It is seen that this tendency is also confirmed in the current phase diagram obtained from the crossover theory. The superfluid phase below $P_{\rm c}$ at $k_{\rm F}a \!<\! 0$ yields the spatial oscillation of the pairing, i.e., the FFLO state. The formation of the FFLO pairing pushes up the phase boundary relative to the Pauli limit (see also Fig.~\ref{fig:TPphase}). 

The behavior of the $P_{\rm c}$ curve in the BEC side is in contrast with that in the BCS side, where the phase boundary is uniquely determined as $P_{\rm c} \!=\! 1$. This results from the fact that the superfluidity survives by locally forming a molecular-like pairing with the corresponding amount of majority spins. The SF phase at $1/k_{\rm F}a \!>\! 0$ corresponds to the PS state without any oscillation of the pairing. The distinct phase boundary between the FFLO and PS states cannot be defined because the FFLO state continuously turns into the PS state via the proximity effect in the BCS/polarized-normal domain interfaces, as has been discussed above. This is peculiar to the finite trap system. 

It has been proposed~\cite{pieri06,pieri07} that as $1/k_{\rm F}a$ reaches the deep BEC limit $1/k_{\rm F}a \!\gg\! 1$, the PS state becomes a ``homogeneous'' imbalanced superfluid without a phase-separated domain, that is, a mixed system of bosons and spinless fermions is attained.

\section{Quasi-particle Structure and Radio-Frequency Spectroscopy}

\begin{figure*}[t!]
\begin{center}
\includegraphics[width=0.7\linewidth]{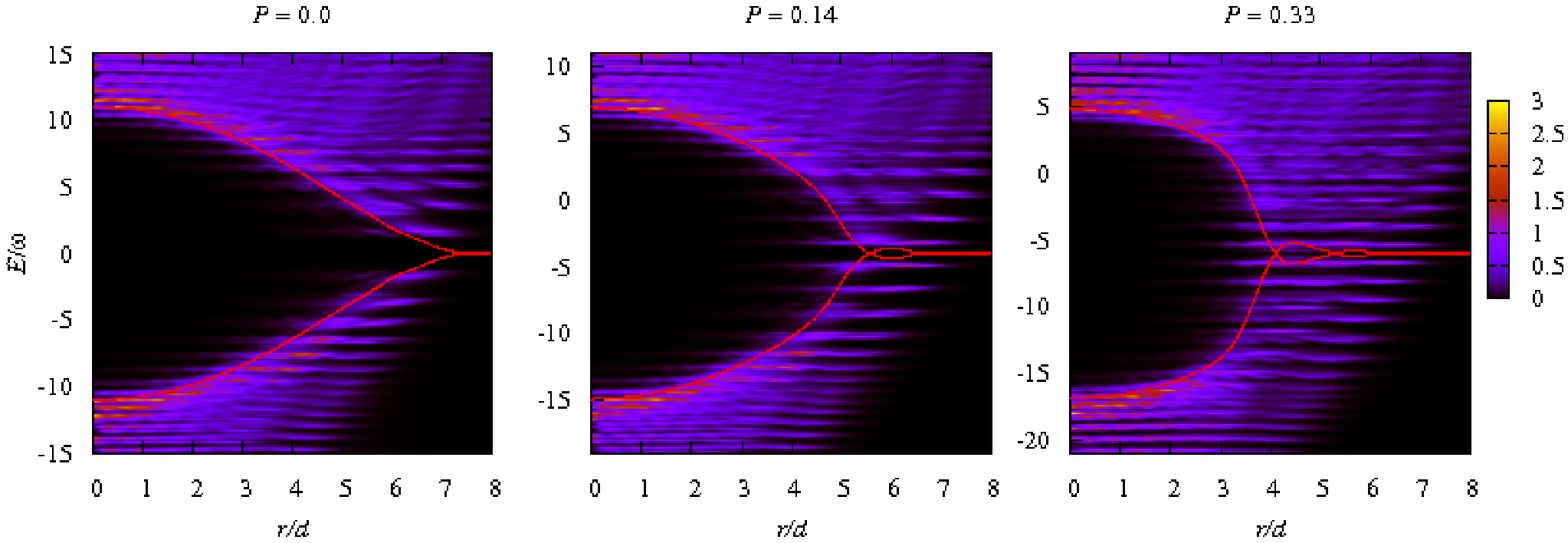} \\
\includegraphics[width=0.7\linewidth]{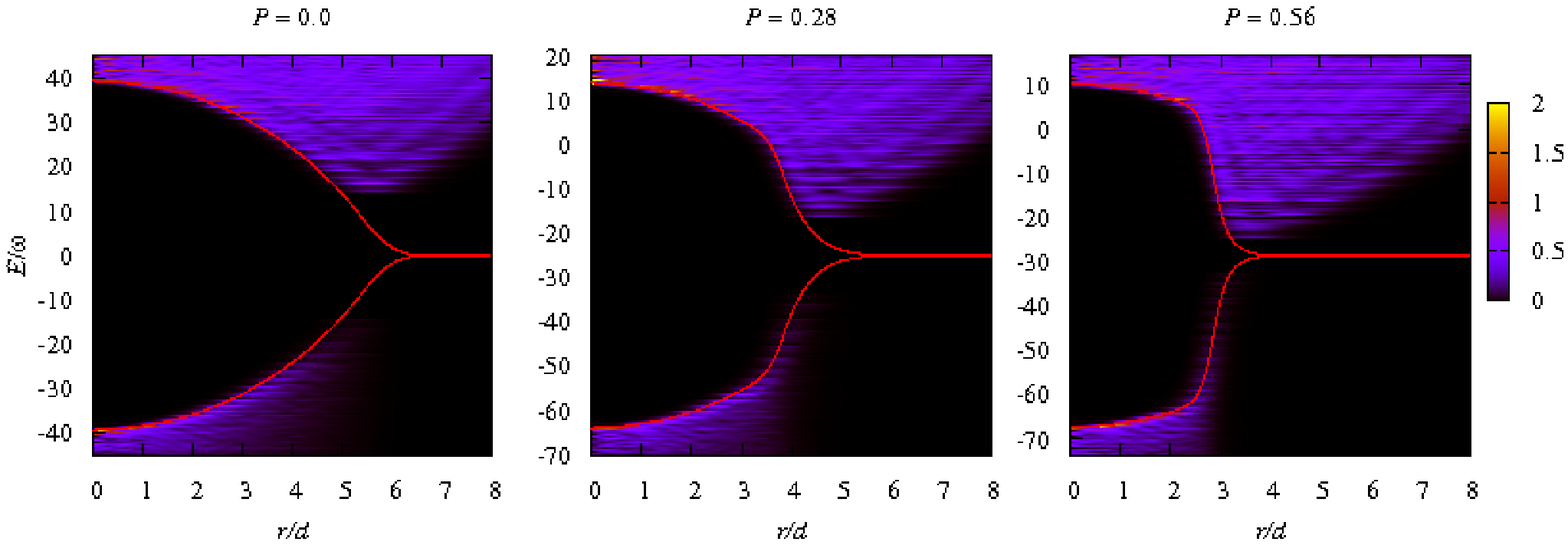}
\end{center}
\caption{
(Color online) LDOS for majority species for various values of $P$. The top row is in the BCS side ($1/k_{\rm F}a \!=\! -0.52$) and the bottom row is in the BEC side ($1/k_{\rm F}a \!=\! +0.52$). The origin of the vertical axis corresponds to the energy equal to the chemical potential $\mu$, and the shift of the gap center from the origin is characterized by the mismatch $\delta \mu$. The solid lines denote the local energy gap defined by $\pm |\Delta(r)|$.
}
\label{fig:ldos}
\end{figure*}

\subsection{Local density of states}

The LDOS for each spin component is given by the definition
\begin{eqnarray}
\mathcal{N}_{\sigma} (r,E) \!=\! - \frac{1}{\pi}
\Im \mathcal{G}_{\sigma}({\bf r}{\bf r},i\omega _n \!\rightarrow\! E + i\eta),
\end{eqnarray} 
where $\mathcal{G}_{\uparrow}({\bf r}{\bf r}',i\omega _n) \!=\! \mathcal{G}_{11}({\bf r}{\bf r}',i\omega _n)$ and $\mathcal{G}_{\downarrow}({\bf r}{\bf r}',i\omega _n) \!=\! -\mathcal{G}_{22}({\bf r}'{\bf r},-i\omega _n)$.
Using the thermal Green's function $\mathcal{G}_{ij}$ described in Appendix B, one can read the LDOS for the spin-up state,
\begin{subequations}
\begin{eqnarray}
\mathcal{N}_{\uparrow} (r,E) = \sum _{\nu} |u_{\nu}(r)|^2 \delta(E- E_{\nu}), 
\end{eqnarray}
and for the spin-down state,
\begin{eqnarray}
\mathcal{N}_{\downarrow} (r,E) = \sum _{\nu} |v_{\nu}(r)|^2 \delta(E+ E_{\nu}).
\end{eqnarray}
\end{subequations}

The LDOS for majority spin states in the BCS side are displayed in the top row of Fig.~\ref{fig:ldos}. In the balanced case, the low-lying excitations are bound in the surface of the cloud ($r \!\ge\! 4d$). It is known~\cite{baranov,OG05,OG052} that the quasi-particles around the surface experience the effective potential consisting of the trap potential $V(r) \!\propto\! r^2$ and the pair potential $\Delta (r)$, where the latter is a monotonically decreasing function in the surface region. This situation may be reduced to a problem on quasi-particles confined in the quantum well like potential $\Delta(r) + V(r)$. Hence, the eigenenergies close to the Fermi level are discretized using the trap unit, i.e., a finite small energy gap of size $\mathcal{O}(\omega)$ exists. For finite $P$, however, the FFLO oscillation induces a mid gap mode with zero energy, and the surface excitation gap vanishes as $P$ increases. The quasi-particles around the FFLO node behave as gapless normal particles, but the superfluidity survives. 

Note that the LDOS profile for the minority species is almost the same as that for the majority spins, except for the shift of the Fermi level, which is shifted downward $2\delta\mu$.

The bottom row in Fig.~\ref{fig:ldos} shows the LDOS in the BEC side. At $1/k_{\rm F}a \!=\! 0.52$, the maximum value of the energy gap is comparable to the Fermi energy $|\Delta _0| \!\sim \! E_{\rm F}$. In contrast, the quasi-particle structure in the surface region is different from that in the BCS side. In the deeper BEC limit, the low-lying excitation can be characterized by the binding energy, that is, $E_{\bf q} \!=\! \sqrt{\mu^2 + \Delta^2} \!\sim\! |\mu|$. Also, the quasi-particle states in energy bands lower than $-\Delta(r) - \delta\mu$ are no longer the eigenstates of the harmonic oscillator, in contrast with those in the weak-coupling region, $1/k_{\rm F}a \!=\! -0.52$. With increasing $P$, as shown in the bottom row of Fig.~\ref{fig:ldos}, the particles, for instance, for $r/d \!>\! 4$ at $P\!=\! 0.56$, locally dissociate from the pairing state, which becomes the normal state.

\subsection{Basic formalism for radio-frequency spectroscopy}

The spatially averaged quantity of the LDOS is observable in the RF spectroscopy, which has been recently developed by several experimental groups, in an equal mixture~\cite{grimm,regalPRL03,gupta,regal03} and an imbalanced system.~\cite{mit4} Theoretical studies have been carried out by a number of authors based on linear response~\cite{bruun01,OG05,OG052,torma,kinnunen04,nygaard, he05} and nonlinear response theories.~\cite{kinnunenPRL06,kinnunen}. Here, following their previous works based on linear response theory, we describe the formulation for RF spectroscopy in an imbalanced system. 

We consider another internal state $| e \rangle$ in addition to the two hyperfine states $| \sigma = \uparrow, \downarrow \rangle$, which form a pairing via an effectively attractive interaction. The state $| e \rangle $ can be described by the field operators ${\psi}_{e} ({\bf r})$ and ${\psi}^{\dag}_{e}({\bf r})$, which obey the standard fermionic commutation relation and are commutative with those of other internal states. On the basis of several works,~\cite{bruun01,OG05,OG052,torma,kinnunen04,nygaard, mahan, he05} we extend the original Hamiltonian describing the state $| \sigma \rangle$ to
\begin{eqnarray}
\tilde{\mathcal{H}} = {\mathcal{H}}_{\rm MF} + {\mathcal{H}}_e + {\mathcal{H}}_{T} 
+ \frac{\omega _{\rm det}}{2}\left[ \sum _{\sigma} \mathcal{N}_{\sigma} - \mathcal{N}_{e} \right] ,
\label{eq:tunH}
\end{eqnarray}
with 
\begin{eqnarray}
\mathcal{N}_{\alpha} \!= \! \int d{\bf r}{\psi}^{\dag}_{\alpha}({\bf r}){\psi}_{\alpha}({\bf r}), \hspace{3mm}
\alpha = \sigma, e .
\end{eqnarray}
The Hamiltonian includes the following contributions. First, the Hamiltonian for atoms in the $e$ state is given by
\begin{eqnarray}
{\mathcal{H}}_e  = \int d{\bf r} 
{\psi}^{\dag}_e({\bf r})
\left[ H^{(0)}_{e} + \sum g_{e\sigma} \rho _{\sigma}({\bf r}) \right]
{\psi}_e({\bf r}),
\end{eqnarray}
with ${\mathcal{H}}_e \!=\! - \frac{\nabla^2}{2M} + V({\bf r}) - \mu _e$. Second, the Hamiltonian describing the ``tunneling current'' between internal states is introduced as 
\begin{eqnarray}
{\mathcal{H}}^{(\sigma)}_{T} = \int  d{\bf r} \left[ 
\Omega ({\bf r}){\psi}^{\dag}_{e}({\bf r}){\psi}_{\sigma}({\bf r}) + {\rm h.c.}
\right].
\label{eq:tunnel1}
\end{eqnarray}
The detuning frequency $\omega _{\rm det}$ expresses the difference between the internal energy level difference and the frequency of the applied laser. 

We calculate the current from the pairing state $| \sigma \rangle$, corresponding to the rate of change of the population of the $e$-state: $I(t) \!=\! \langle \dot{\mathcal{N}}_e (t)\rangle $. Here we divide the Hamiltonian in eq.~(\ref{eq:tunH}) into two parts: (i) the diagonal part $\tilde{\mathcal{H}}_0 \!=\! {\mathcal{H}}_{\rm MF} + {\mathcal{H}}_e + \frac{\omega _{\rm det}}{2}[ \sum _{\sigma} \mathcal{N}_{\sigma} - \mathcal{N}_{e} ]$ and (ii) the perturbation Hamiltonian $\mathcal{H}_T$. From linear response theory~\cite{mahan}, the tunneling current is given as 
\begin{eqnarray}
I^{(\sigma)}(t) = - i\int dt' \theta(t-t') \langle [\dot{\mathcal{N}}_e(t), {\mathcal{H}}^{(\sigma)}_T (t')] \rangle ,
\label{eq:kubo}
\end{eqnarray}
where the expression of the time-dependent quantity is given by $\mathcal{O}(t) \!=\! {\rm e}^{i\tilde{H}'_0 t}\mathcal{O}{\rm e}^{-i\tilde{H}'_0t}$ with the canonical Hamiltonian $\tilde{\mathcal{H}}'_0 \!=\! \tilde{\mathcal{H}}_0 + \mu _e \mathcal{N}_e + \sum _{\sigma}\mu _{\sigma}\mathcal{N}_{\sigma}$. In particular, the tunneling Hamiltonian defined in eq.~(\ref{eq:tunnel1}) is transformed to
\begin{eqnarray}
{\mathcal{H}}^{(\sigma)}_T (t) = \int d{\bf r} \Omega({\bf r})\left[ {\rm e}^{-i\tilde{\omega}t}\psi^{\dag}_{e}({\bf r},t)\psi _{\sigma}({\bf r},t) + \mbox{h.c.} \right],
\label{eq:tunnel2}
\end{eqnarray}
using the field operators in the Heisenberg representation, $\psi _{e}({\bf r},t) \!=\! {\rm e}^{i\mathcal{H}_et}\psi _{e}({\bf r}){\rm e}^{-i\mathcal{H}_et}$ and $\psi _{\sigma}({\bf r},t) \!=\! {\rm e}^{i\mathcal{H}_{\rm MF}t}\psi _{\sigma}({\bf r}){\rm e}^{-i\mathcal{H}_{\rm MF}t}$. Hereafter, we use the notation $\tilde{\omega} \!\equiv\! \omega _{\rm det} + \mu _{\sigma} - \mu _e$ for convenience. Also, the number operator for the $e$-state obeys the Heisenberg equation, $i\dot{\mathcal{N}}_e (t) = [\mathcal{N}_e(t), \tilde{\mathcal{H}}]$, which leads to the following expression for the Heisenberg representation,
\begin{eqnarray}
\dot{\mathcal{N}}_e (t) = -i \int d{\bf r} \Omega({\bf r})
\left[ {\rm e}^{-i\tilde{\omega} t} \psi^{\dag}_e ({\bf r},t)\psi _{\sigma}({\bf r},t) - \mbox{h.c.} \right].
\label{eq:enum}
\end{eqnarray} 

By substituting Eqs.~(\ref{eq:tunnel2}) and (\ref{eq:enum}) into the Kubo formula in eq.~(\ref{eq:kubo}), one can see that the total current is composed of two contributions, a single-particle tunneling current $I^{(\sigma)}_S$ and a Josephson tunneling current $I^{(\sigma)}_J$, as $I^{(\sigma)}\!\equiv\!I^{(\sigma)}_S+I^{(\sigma)}_J$. Since we are interested in single-particle tunneling, the Josephson current is neglected here. The resulting single-particle tunneling current is obtained from the analytic continuation of the quantity expressed as the product of the Green's functions for the $\sigma$- and $e$-states~\cite{mahan}: 
\begin{subequations}
\begin{eqnarray}
I^{(\sigma)}_S(t) = 2 \Im \int\int  \mathcal{U}_{\sigma}({\bf r}{\bf r}',i\omega _n \rightarrow \tilde{\omega}+i\eta)
d{\bf r} d{\bf r}',
\end{eqnarray}
where the Matsubara frequency is introduced as $\omega _n \!\equiv\! \pi(2n + 1)\beta$ and 
\begin{eqnarray}
\hspace{-3mm}\mathcal{U}_{\uparrow}({\bf r}{\bf r}',i\omega _n ) 
= \beta^{-1}\Omega^{\ast}({\bf r})\Omega({\bf r}')\sum _{\omega '_n} \hspace{12mm} \nn \\
\times \mathcal{G}_{11}({\bf r}'{\bf r},i\omega '_n-i\omega _n) \mathcal{G}_{e}({\bf r}{\bf r}',i\omega '_n) , 
\end{eqnarray}
\begin{eqnarray}
\hspace{-3mm}\mathcal{U}_{\downarrow}({\bf r}{\bf r}',i\omega _n ) 
= \beta^{-1}\Omega^{\ast}({\bf r})\Omega({\bf r}')\sum _{\omega '_n} \hspace{12mm} \nn \\
\times \mathcal{G}_{22}({\bf r}{\bf r}',i\omega _n-i\omega '_n) \mathcal{G}_{e}({\bf r}{\bf r}',i\omega '_n) .
\end{eqnarray}
\end{subequations}
The Green's function for the $e$-state is given as 
\begin{eqnarray}
\mathcal{G}_{e}({\bf r}{\bf r}',i\omega _n) 
= \sum _{\zeta} \frac{\phi _{\zeta}({\bf r})\phi^{\ast}_{\zeta}({\bf r})}{i\omega _n - \epsilon _{\zeta}},
\end{eqnarray}
where the eigenfunction and energy, $\phi _{\zeta}$ and $\epsilon _{\zeta}$, are obtained from the Schr\"{o}dinger equation for atoms in the $e$-state, $[H^{(0)}_e + \sum g_{e\sigma}\rho _{\sigma}]\phi _{\zeta}({\bf r}) \!=\! \epsilon _{\zeta}\phi _{\zeta}({\bf r})$. The expressions of the Matsubara Green's function for the pairing state $\mathcal{G}_{11}$ and $\mathcal{G}_{22}$ are shown in Appendix B. 

To this end, one can find single-particle tunneling currents for the following distinguishable processes: for the tunneling from the majority species $| \sigma = \uparrow \rangle$ to the $e$-state,
\begin{subequations}
\label{eq:current}
\begin{eqnarray}
I^{(\uparrow)}_{S} (\tilde{\omega}) = 2\pi \sum_{\nu , \zeta} 
\left |\int \Omega({\bf r})u_{\nu}({\bf r})\phi^{\ast}_{\zeta}({\bf r}) d{\bf r} \right|^2 \hspace{10mm} \nn \\
\times [f_{\nu}-f(\epsilon _{\zeta})] \delta(\tilde{\omega}+E_{\nu}-\epsilon _{\zeta}),
\label{eq:current1}
\end{eqnarray}
and for the tunneling from the minority species $| \sigma = \downarrow \rangle$ to the $e$-state, 
\begin{eqnarray}
I^{(\downarrow)}_{S} (\tilde{\omega}) = 2\pi \sum_{\nu , \zeta} 
\left |\int \Omega({\bf r})v^{\ast}_{\nu}({\bf r})\phi^{\ast}_{\zeta}({\bf r}) d{\bf r} \right|^2 \hspace{10mm} \nn \\
\times [f_{\nu}-f(\epsilon _{\zeta})] \delta(\tilde{\omega}-E_{\nu}-\epsilon _{\zeta}).
\label{eq:current2}
\end{eqnarray}
\end{subequations}
Note that similarly to the BdG formalism in \S~2, the summation in Eqs.~(\ref{eq:current1}) and (\ref{eq:current2}) is carried out for all eigenstates with both positive and negative energies because of the breaking of the time-reversal symmetry. 

In performing the numerical calculation, the $\delta$-function in eq.~(\ref{eq:current}) is replaced with the Lorentzian function $\delta (z) \!\rightarrow\! \Gamma _{\eta} (z) \!=\! (\eta/2)^2/[z^2+(\eta/2)^2]$, where the resolution of the spectrum $\eta$ is set as $\eta \!=\! 1.0\omega$ throughout this paper. We consider the situation when the interaction between the pairing state and the $e$-state is negligible, $g_{e\sigma} \!=\! 0$. We also focus on the transition from the pairing state $|\sigma\rangle$ to the excited state $|e\rangle$, i.e., the positive detuning $\omega _{\rm det} \!>\! 0$ and $I^{(\sigma)}_S \!>\! 0$. 

\subsection{Numerical results}

It is important to mention that the tunneling current in the homogeneous pairing field $\Delta ({\bf r}) \!=\! \Delta$ at $P\!=\! 0$ is expressed as
\begin{eqnarray}
I^{(\sigma)}_S(\omega _{\rm det}) = -  \sum _{{\bf p},{\bf q}} \left|\Omega _{{\bf p}{\bf q}}\right|^2 
\int^{\infty}_{-\infty} \frac{d\xi}{2\pi}
[ f(\xi - \mu _e + \omega _{\rm det}) \hspace{5mm} \nn \\ 
- f(\xi - \mu _e) ]
\mathcal{N}_{\sigma}({\bf q},\xi - \mu _{\sigma}) \mathcal{N}_{e}({\bf p},\xi + \omega _{\rm det} - \mu _{e}) ,
\label{eq:current_homo}
\end{eqnarray}
with $\Omega _{{\bf p}{\bf q}} \!\equiv\! \int d{\bf r} {\rm e}^{i{\bf p}\cdot{\bf r}}\Omega({\bf r}){\rm e}^{-i{\bf q}\cdot{\bf r}}$. The density of states for the pairing state is 
\begin{eqnarray}
\mathcal{N}_{\sigma}({\bf q},z) = 2\pi 
\left[ u^2_{\bf q} \delta(z - E_{\bf q}) + v^2_q \delta(z + E_{\bf q}) \right] ,
\end{eqnarray}
and that for the $e$-state is $\mathcal{N}_e (z) \!=\! 2\pi \delta(z - \epsilon _{\bf p})$. Here, $(u_{\bf q}, v_{\bf q})$ and $E_{\bf q}$ are the solutions of the BdG equation in the homogeneous system at $P\!=\! 0$, and $\epsilon _{\bf p} \!=\! p^2/2m - \mu _e$ is the eigenenergy of a free particle in the $e$-state. We are interested in the situation of positive detuning and current, $\omega _{\rm det} \!>\! 0$ and $I^{(\sigma)}_S \!>\! 0$, which expresses the fraction loss of the pairing $\sigma$-state. First, one obtains the following simple result in the case of $\Delta \!=\! 0$,
\begin{eqnarray}
I^{(\sigma)}_S(\omega _{\rm det}) = 2\pi \Omega^2 \left( N_{\sigma} - N_{e} \right)\delta(\omega _{\rm det}),
\end{eqnarray}
which leads to a single-peak structure at $\omega _{\rm det} \!=\! 0$ when the $e$-state is initially not occupied ($\mu _e \!=\! 0$). The intensity of the peak gradually decreases with increasing $\mu _e$, and in the case of the equal chemical potential $\mu _e \!=\! \mu _{\sigma}$, the tunneling current is not responsible to any detuning frequencies, i.e., $I^{(\sigma)}_S \!=\! 0$. 

For the pairing state $\Delta \!\neq\! 0 $, after evaluating the integral in eq.~(\ref{eq:current_homo}) over energy and momenta, one obtains~\cite{torma}
\begin{eqnarray}
I^{(\sigma)}_S (\omega _{\rm det}) = \pi \Omega^2 
\mathcal{N}_0\left( \frac{\omega^2_{\rm det}-\Delta^2}{\omega _{\rm det}} + 2\mu _{\sigma} \right) 
\frac{\Delta^2}{\omega^2_{\rm det}} \nn \\
\times \Theta\left( \omega^2_{\rm det} - \Delta^2 + 2\delta\tilde{\mu} \omega _{\rm det} \right), \hspace{10mm}
\label{eq:current_solution}
\end{eqnarray}
where we introduce the density of states in the ideal Fermi gas, $\mathcal{N}_0(E) \!=\! \frac{V}{2\pi^2}\sqrt{E}$. Also, we set $\delta\tilde{\mu} \!\equiv\! \mu _{\sigma} - \mu _e$. For positive detuning $\omega _{\rm det} \!>\! 0$ and a system with positive chemical potential $\mu _{\sigma} \!>\! 0$, it is found that the function $I^{(\sigma)}_S(\omega _{\rm det})$ becomes monotonically decreasing in the range $\omega _{\rm det}/E_{\rm F} \!>\! \Delta /E_{\rm F}$ when the $e$-state is initially occupied, $\mu _e \!=\! \mu _{\sigma}$. In contrast, in the case of $\mu _{e} \!=\! 0$, the resonant detuning at which $I^{(\sigma)}_S$ has the maximum value is shifted to $\omega _{\rm det}/E_{\rm F} \!=\!\frac{5}{8}(\frac{\Delta}{E_{\rm F}})^2$. In the deep BEC limit where $\mu _{\sigma}\!<\! 0$ and $\Delta/|\mu _{\sigma}| \!\ll\! 1$, $I^{(\sigma)}_S$ exhibits different behavior from that in the BCS side, which is insensitive to $\mu _e$. Then, the resonant detuning is situated around $\omega _{\rm det}/E_{\rm F} \!\sim\! 2|\mu _{\sigma}|/E_{\rm F} \!\sim\! |E_b|/E_{\rm F}$. This energy corresponds to the dissociation of molecular bosons and is uniquely characterized by the dimensionless parameter $1/(k_{\rm F}a)^2$.

\begin{figure}[b!]
\begin{center}
\includegraphics[width=0.9\linewidth]{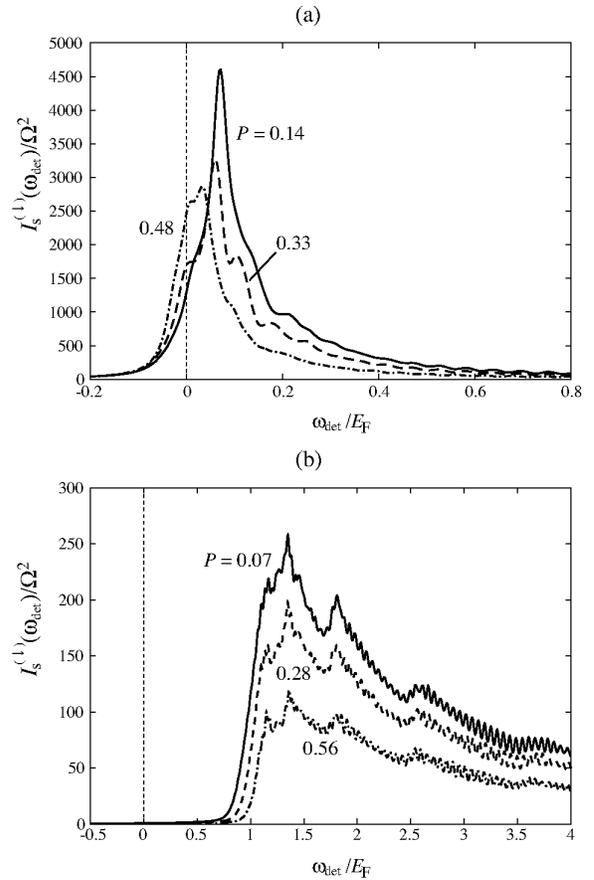}
\end{center}
\caption{
RF spectroscopy of the minority component $I^{(\downarrow)}_S$ at (a) $1/k_{\rm F}a \!=\! -0.52$ and (b) $1/k_{\rm F}a \!=\! 0.52$ with $N\!=\! 3,000$ atoms. All the results are at $T\!=\! 0$.
}
\label{fig:T0current}
\end{figure}

In Fig.~\ref{fig:T0current}, we display the numerical results of the fraction loss of the minority species $I^{(\downarrow)}_S$ in imbalanced systems in the presence of a harmonic trap, where (a) and (b) correspond to the BCS and BEC sides, respectively. Hereafter, we consider the situation that the $e$-state is initially empty, $\mu _e \!=\! 0$. Also, we set $\Omega({\bf r}) \!=\! \Omega$. In the BCS side, as shown in Fig.~\ref{fig:T0current}(a), the resonant detuning for low values of $P$ is situated around $\omega _{\rm det}/E_{\rm F} \!\simeq\! 0.1$, which is related to the dissociation energy of the fermionic pairing described above, $\omega _{\rm det}/E_{\rm F} \!\sim\! (\Delta _0/E_{\rm F})^2$, with the maximum gap of $\Delta _0 \!=\! 0.34 E_{\rm F}\!=\! 11\omega$. As $P$ increases, the peak position approaches the zero detuning. Note that for high values of $P$, additional fraction loss occurs at the zero detuning and the resulting spectrum profile has a double-peak structure. This reveals the fact that the system under high imbalance is in the pairing state with partially polarized spins, which is indirect evidence of the FFLO state. As seen in Fig.~\ref{fig:ldos}, the mid gap state appears in the spacing between the small energy gap near the surface ($r/d \!\sim\!6$) when $P \!\neq\! 0$. The presence of the mid gap state increases the intensity of the fraction loss at $\omega _{\rm det} \!=\! 0$.

It is seen from Fig.~\ref{fig:T0current}(b) that the fraction loss in the BEC side yields a spectrum distinctive from that in the BCS side. There are two differences. (i) The detuning at which the fermionic/molecular pairs are dissociated is unchanged for increasing values of $P$. Only the intensity becomes weak. (ii) With increasing $P$, no additional peak appears around the zero detuning. This is because all the minority spins in the imbalanced situation form the ``pairs'' with the corresponding amount of the majority species in the local region, and the spins in the normal state are fully polarized. The fraction loss for the majority species is unchanged in the extensive region from the BCS limit to the BEC limit, where the spectrum always yields the double-peak structure with one peak situated at $\omega _{\rm det} \!=\! 0$ having large intensity and the other depending on the dissociation energy of fermionic or molecular pairs at $\omega _{\rm det} \!\simeq\! \Delta _0$. 
 
\begin{figure}[t!]
\begin{center}
\includegraphics[width=0.9\linewidth]{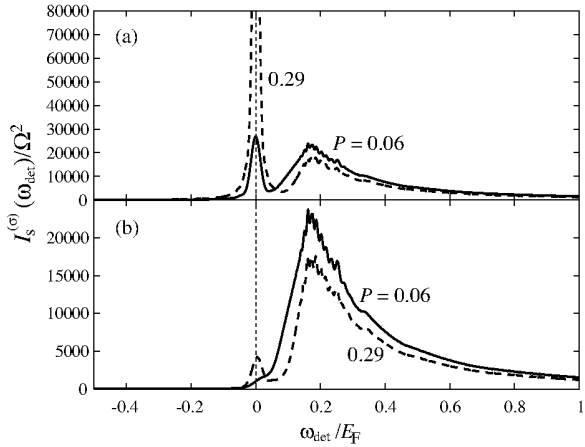}
\end{center}
\caption{
Fraction loss of (a) majority and (b) minority species at $1/k_{\rm F}a \!=\! -0.14$ with $N\!=\! 150,000$ atoms. All the results are at $T\!=\! 0$.
}
\label{fig:T0current2}
\end{figure}

Finally, the RF spectroscopy for the system with a realistic number of particles $N\!=\! 150,000$ in the vicinity of the unitary limit $1/k_{\rm F}a \!=\! -0.14$ is presented in Fig.~\ref{fig:T0current2}. The qualitative behavior is unchanged from the case of the small particle number displayed in Fig.~\ref{fig:T0current}, that is, the fraction loss of the minority species occurs at zero detuning in addition to $\omega _{\rm det}/E_{\rm F} \!\sim\! \frac{5}{8}(\frac{\Delta _0}{E_{\rm F}})^2 \!=\! 0.17$, corresponding to the dissociation energy of the pairing state at $r \!=\! 0$, $\Delta _0 \!=\! 0.53 E_{\rm F} \!=\! 82 \omega$. The corresponding pairing field is shown in Fig.~\ref{fig:largeNg-5}, where the FFLO modulation appears in the vicinity of the boundary between the equal-pairing core and polarized normal domain. It should be emphasized that the satellite peak at zero detuning is indirect evidence for the FFLO pairing. This prediction can be experimentally checked by carefully examining the RF spectroscopy in the lower temperature region.~\cite{mit4} Also note that the spectrum presented in Fig.~\ref{fig:T0current2} is in good agreement with that obtained from nonlinear response theory.~\cite{kinnunen}

\section{Concluding Remarks}

In this paper, we have theoretically studied the stable superfluid state in strongly interacting trapped Fermi systems with population imbalance, based on the the single-channel Hamiltonian. We have numerically solved the BdG equation coupled with the regularized gap equation and the number equation in the BCS-BEC crossover regime under imbalanced spin densities, where the computation for the higher-energy contribution was supplemented by the LDA. 

The main results are twofold: (1) First, in \S~3, we have discussed the ground state in the crossover regime under population imbalance and presented the quantum phase diagram in the $1/k_{\rm F}a$-$P$ plane. In the weak-coupling regime ($1/k_{\rm F}a \!<\! 0$), it has been found that the resulting pairing field at $T=0$ exhibits the FFLO oscillation around the edge of the minority component. In particular, the pairing field exhibits the oscillation in the entire region of the system when $P$ approaches $P_{\rm c}$. This novel pairing state is reflected in the density and local magnetization profiles, such as the bimodal structure for the minority species. In contrast, the FFLO oscillation disappears for all $P$ in the BEC regime, where the resulting ground state yields the phase separation between the balanced pairing domain and the fully polarized spin domain. We have found that the spatial variation of the pairing field affects the density. For instance, the presence of the FFLO-modulated pairing field leads to a partially polarized spin density, while the PS state is reflected in the fully polarized spins. It has also been shown that the FFLO modulation survives even in the unitary limit as the proximity effect. We have confirmed that these tendencies of the ground-state structure are unchanged in a system with a realistic particle number $N \!\sim\! \mathcal{O}(10^5)$, which is comparable with recent experiments.~\cite{rice1,rice2,mit1,mit2,mit3,mit4} Our calculations reproduce a PS-like profile in the local magnetization, while the pairing field yields the FFLO modulation even in the vicinity of the resonance. The periodicity and intensity of the modulation increase as $1/k_{\rm F}a$ approaches the weak coupling BCS regime. In particular, we have found that the periodicity $L$ of the FFLO oscillation is well scaled with the coherence length $\xi _0$ as $L \!\sim\! 3\xi _0$, throughout the extensive range of $k_{\rm F}a \!\le\! 0$ and $P \!<\! P_{\rm c}$.

(2) The second part of the present paper has been devoted to another observable quantity, the RF spectroscopy. By numerically solving the tunneling current derived from linear response theory, the contributions of the pairing field to the spectrum have been discussed. The clear difference in the resonance shape between the BCS and BEC sides reveals the different superfluid state, which can be checked by further experiments. 

Note that in the phase contrast imaging of the local magnetization by the MIT experiment\cite{mit3}, a PS-like profile was observed at the unitary limit $1/k_{\rm F}a \!=\! 0$. Our results presented in Figs.~\ref{fig:gapT0uni} and \ref{fig:largeNg-5} are in good agreement with this profile, which implies that the pairing field exhibits the FFLO oscillation at the edge of the cloud. Also, we have demonstrated that as the system approaches the weak coupling BCS regime, the FFLO modulation covers the entire region of the system, particularly at $P \!\sim\! P_{\rm c}$. In a previous experiment,\cite{mit1} the quantum phase transition in the BCS side of the resonance, $1/k_{\rm F}a \!\sim\! -0.4$, has already been observed, which is a favorable condition for the detection of the FFLO. Further detailed analysis may catch the signature of the FFLO pairing via the density and RF spectroscopy results, as we have described in the present paper. 

\begin{figure}[t!]
\begin{center}
\includegraphics[width=0.9\linewidth]{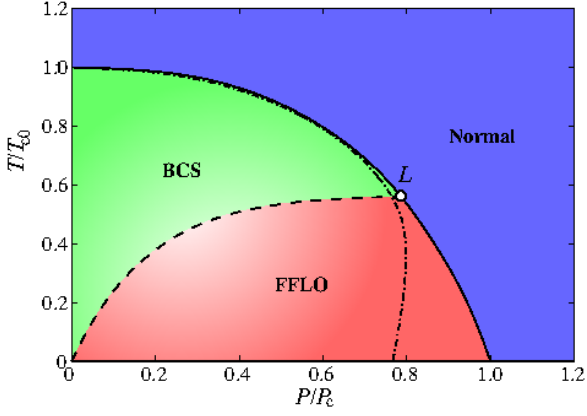} 
\end{center}
\caption{
(Color online) Typical $T$-$P$ phase diagram in the weak coupling BCS side of a resonance ($1/k_{\rm F}a \!=\! -0.75$). The dashed-dotted line denotes $T_{\rm c}$ for the non-oscillating BCS state.~\cite{MMI3} Empty circle is the Lifshitz point. 
}
\label{fig:TPphase}
\end{figure}

Finally, we comment on the thermodynamic stability of the FFLO phase against the increase in $T$. In Fig.~\ref{fig:TPphase}, we present a phase diagram in the $T$-$P$ plane in the BCS side. The phase diagram yields multiple superfluid phases composed of the FFLO state and the nonoscillating BCS state. All three phase transition lines between the BCS, FFLO, and normal states are of the second order, and these lines meet at the so-called Lifshitz ($L$) point.~\cite{chaikin} The BCS-FFLO line starts from $P \!=\!T \!=\! 0$, implying that the ground state is always the FFLO state when $P \!\neq\! 0$ and $T$ is low. The BCS state only appears at higher values of $T$. This second-order phase transition via the FFLO state may be realized in the weak-coupling regime, while in the strong-coupling regime the first-order transition is predicted by a LDA calculation~\cite{gubbels} in which the Lifshitz point is replaced by the tricritical point. We also note that this $L$ point exhibits the universal temperature $T_{L}/T_{c0} \!\simeq\! 0.6$ in the weak-coupling regime, independent of the coupling constant $1/k_{\rm F}a$. Hence, all three second-order lines are uniquely determined with a fixed $L$ point. A similar phase diagram has been proposed even in the absence of the trap potential.~\cite{he2}

One of the main outcomes in the current work is that the FFLO-modulated pairing field survives in the whole region in the system approaching the weak-coupling limit ($1/k_{\rm F}a \!\rightarrow\! -\infty$). The calculations presented here have been performed in a system restricted to cylindrical geometry. The FFLO oscillating pattern in a fully three-dimensional system without any restriction, such as an elongated cigar-shaped or disk-shaped trap, is still open to question, and should be further explored in future. 

\acknowledgements

The authors acknowledge the support of a Grant-in-Aid for Scientific Research from the Japan Society for the Promotion of Science.

\appendix

\section{Regularized BdG Equation}

We start with the original Hamiltonian in eq.~(\ref{eq:original}). Here, it is convenient to introduce a spinor in the Nambu space,
\begin{eqnarray}
\mbox{\boldmath $\Psi$}({\bf r}) = [{\psi}_{\uparrow}({\bf r}),{\psi}^{\dag}_{\downarrow}({\bf r})]^{T}.
\end{eqnarray}
Applying the standard mean-field approximation to the interaction part of the above Hamiltonian, the effective Hamiltonian can be derived as
\begin{eqnarray}
{\mathcal{H}}_{\rm MF} = \mathcal{E}_0 + \int d{\bf r} \int d{\bf r}' \mbox{\boldmath $\Psi$}^{\dag}({\bf r}) \hat{\mathcal{K}}({\bf r},{\bf r}')\mbox{\boldmath $\Psi$}({\bf r}'),
\label{eq:mfH}
\end{eqnarray}
where 
\begin{subequations}
\begin{eqnarray}
\hat{\mathcal{K}}({\bf r},{\bf r}') \equiv   
\left[
\begin{array}{cc}
\mathcal{K}_{\uparrow}({\bf r},{\bf r}') & \Delta({\bf r},{\bf r}') \\
\Delta^{\ast}({\bf r},{\bf r}') &  - \mathcal{K}^{\ast}_{\downarrow}({\bf r},{\bf r}')
\end{array}
\right], 
\label{eq:matrix}
\end{eqnarray}
\begin{eqnarray}
\mathcal{K}_{\sigma}({\bf r},{\bf r}') = H^{(0)}_{\sigma}\delta({\bf r}-{\bf r}') + \mathcal{W}_{-\sigma}({\bf r},{\bf r}').
\label{eq:matrix2}
\end{eqnarray}
\end{subequations}
The mean-field quantities, the pairing potential $\Delta({\bf r},{\bf r}')$ and the Hartree potential $\mathcal{W} _{\sigma}({\bf r},{\bf r}')$, are defined by
\begin{subequations}
\begin{eqnarray}
\Delta({\bf r},{\bf r}') \equiv  U(\tilde{r}) \langle {\psi}_{\downarrow}({\bf r}') 
{\psi}_{\uparrow}({\bf r})\rangle, 
\label{eq:delta}
\end{eqnarray}
\begin{eqnarray}
\mathcal{W}_{\sigma}({\bf r},{\bf r}') \equiv U(\tilde{r})\rho _{\sigma}({\bf r}) 
= U(\tilde{r}) \langle {\psi}^{\dag}_{\sigma}({\bf r})	{\psi}_{\sigma}({\bf r})\rangle.
\label{eq:hartree}
\end{eqnarray}
\end{subequations}
$\mathcal{E}_0$ is the $c$-number including the condensation and Hartree energies and $\tilde{r} \!\equiv\! | \tilde{\bf r} | \!=\! |{\bf r} - {\bf r}'|$ is the relative coordinate.

The Bogoliubov transformation of the spinor $\mbox{\boldmath $\Psi$}({\bf r})$ into the quasi-particle basis $\mbox{\boldmath $\eta$}_{\nu} \!\equiv\! [{\eta}_{\nu,\uparrow},{\eta}^{\dag}_{\nu , \downarrow}]^T$ is defined by
\begin{eqnarray}
\mbox{\boldmath $\Psi$}({\bf r}) = \sum _{\nu} 
\left[ \begin{array}{cc}
u_{\nu}({\bf r}) & -v^{\ast}_{\nu}({\bf r}) \\ v_{\nu}({\bf r}) & u^{\ast}_{\nu}({\bf r}) 
\end{array} \right]
\mbox{\boldmath $\eta$}_{\nu} \equiv \sum _{\nu} \hat{u}_{\nu} ({\bf r})\mbox{\boldmath $\eta$}_{\nu}.
\label{eq:bogo}
\end{eqnarray}
Here, the creation and annihilation operators of the quasi-particles, ${\eta}^{\dag}_{\nu,\sigma}$ and ${\eta}_{\nu,\sigma}$, obey the fermionic commutation relations. The quasi-particle wave function in the matrix form $\hat{u}_{\nu}({\bf r})$ satisfies the orthonormal condition,
\begin{eqnarray}
\int d{\bf r} \hat{u}^{\dag}_{\nu}({\bf r})\hat{u}_{\nu'}({\bf r}) = \delta _{\nu,\nu'}.
\end{eqnarray}
Also, we write the completeness in a matrix form,
\begin{eqnarray}
\sum _{\nu} \hat{u}_{\nu}({\bf r}) \hat{u}^{\dag}_{\nu}({\bf r}') = \delta({\bf r}-{\bf r}').
\end{eqnarray}
Then, we assume that the mean-field Hamiltonian can be transformed into the diagonalized form, ${\mathcal{H}}_{\rm MF} \!=\! \mathcal{E}_0 + \sum _{\nu}\sum _{\sigma} \varepsilon^{(\sigma)}_{\nu}{\eta}^{\dag}_{\nu,\sigma}{\eta}_{\nu,\sigma}$. To this end, we obtain the BdG equation
\begin{eqnarray}
\int d{\bf r}' \hat{\mathcal{K}}({\bf r},{\bf r}')\hat{u}_{\nu}({\bf r}')
= \hat{u}_{\nu}({\bf r})
\left[\begin{array}{cc}
\varepsilon ^{(\uparrow)}_{\nu} & 0 \\ 0 & - \varepsilon^{(\downarrow)}_{\nu}
\end{array} \right].
\label{eq:bdg}
\end{eqnarray}
The above BdG matrix $\hat{\mathcal{K}}$ yields double eigenstates for hyperfine spins. To see this, we set the eigenfunction with the up spin as $\mbox{\boldmath $\varphi$}^{(\uparrow)}_{\nu} \!\equiv\! [u_{\nu},v_{\nu}]^{T}$, with an eigenvalue of $\varepsilon^{(\uparrow)} _{\nu}$: $\hat{\mathcal{K}}\mbox{\boldmath $\varphi$}^{(\uparrow)}_{\nu} \!=\! \varepsilon^{(\uparrow)}_{\nu}\mbox{\boldmath $\varphi$}^{(\uparrow)}_{\nu}$. It is found that the BdG equation (\ref{eq:bdg}) simultaneously has eigenstates for down spins of $\mbox{\boldmath $\varphi$}^{(\downarrow)}_{\nu}\!\equiv\! [-v^{\ast}_{\nu},u^{\ast}_{\nu}]^{T}$ with the eigenvalue $-\varepsilon^{(\downarrow)}_{\nu}$. Therefore, one can solve the BdG equation 
\begin{eqnarray}
\int d{\bf r}' \hat{\mathcal{K}}({\bf r},{\bf r}')\mbox{\boldmath $\varphi$}_{\nu}({\bf r}')
=  E_{\nu}\mbox{\boldmath $\varphi$}_{\nu}({\bf r}),
\label{eq:bdg2}
\end{eqnarray}
using the eigenfunction $\mbox{\boldmath $\varphi$}_{\nu} \!\equiv\! [u_{\nu},v_{\nu}]^{T}$ and the eigenstates $E_{\nu}$ corresponding to those for up and down spins. However, we should emphasize that $\varepsilon^{(\uparrow)}_{\nu}\neq \varepsilon^{(\downarrow)}_{\nu}$ because the finite mismatch of the Fermi surface causes the breaking of the time-reversal symmetry, $\hat{\mathcal{K}}({\bf r},{\bf r}') \!\neq\! - \hat{\tau}_2\hat{\mathcal{K}}^{\ast}({\bf r},{\bf r}')\hat{\tau}_2$, where $\hat{\tau}_j$ is the $j$th Pauli matrix.

Using the eigenstates in eq.~(\ref{eq:bdg2}), the pairing field defined in eq.~(\ref{eq:delta}) can be explicitly expressed as
\begin{eqnarray}
\Delta({\bf r},{\bf r}') =  U(\tilde{r}) \sum _{\nu}
\left[ u_{\nu}({\bf r})v^{\ast}_{\nu}({\bf r}')f^{(\uparrow)}_{\nu} - v^{\ast}_{\nu}({\bf r})u_{\nu}({\bf r}')f^{(\downarrow)}_{\nu} \right] \nn \\
= U(\tilde{r}) \sum _{\nu}
u_{\nu}({\bf r})v^{\ast}_{\nu}({\bf r}')f_{\nu}, \hspace{30mm}
\label{eq:gap_nsym}
\end{eqnarray}
for the Fermi distribution functions $f^{(\uparrow)}_{\nu} \!\equiv\! f(\varepsilon^{(\uparrow)}_{\nu}) \!=\! \langle {\eta}^{\dag}_{\nu',\uparrow} {\eta}_{\nu,\uparrow}\rangle \delta _{\nu,\nu'}$, $f^{(\downarrow)}_{\nu} \!\equiv\! f(-\varepsilon^{(\downarrow)}_{\nu}) \!=\! \langle {\eta}_{\nu',\downarrow} {\eta}^{\dag}_{\nu,\downarrow}\rangle \delta _{\nu,\nu'}$, and $f_{\nu} \!\equiv\! f(E_{\nu}) \!=\! 1/ ({\rm e}^{E_{\nu}/T}+1)$. The summation in the gap equation (\ref{eq:gap_nsym}) is performed for all the eigenstates with both positive and negative eigenenergies. 
Then, the particle density in each spin state is given from Eqs.~(\ref{eq:hartree}) and (\ref{eq:bogo}), and is described in eq.~(\ref{eq:defrho}). The BdG equation (\ref{eq:bdg2}) is now self-consistently solved under the mean-field conditions, given by Eqs.~(\ref{eq:gap_nsym}) and (\ref{eq:defrho}), for the fixed total particle number:
$N=\sum _{\sigma}N_{\sigma}= \int d{\bf r} \sum _{\sigma}\rho _{\sigma}({\bf r})$.

Now, let us derive the explicit expression for an equal mixture of two-component fermions, i.e., $\mathcal{K}_{\uparrow}\!=\!\mathcal{K}_{\downarrow}$ under $\delta\mu\!=\!0$. Then, the BdG matrix (\ref{eq:matrix}) yields the time-reversal symmetry, $\hat{\mathcal{K}}({\bf r},{\bf r}') \!=\! - \hat{\tau}_2\hat{\mathcal{K}}^{\ast}({\bf r},{\bf r}')\hat{\tau}_2$. It hence follows that a positive eigenvalue $E_{\nu}$ having the eigenfunction $\mbox{\boldmath $\varphi$}_{\nu} \!\equiv\! [u_{\nu},v_{\nu}]^{T}$ in eq.~(\ref{eq:bdg2}) is always accompanied by the negative eigenvalue $-E_{\nu}$ with the eigenfunction $-i\hat{\tau}_2\mbox{\boldmath $\varphi$}^{\ast}_{\nu} \!\equiv\! [-v^{\ast}_{\nu},u^{\ast}_{\nu}]^{T}$. Then, the mean-field quantities shown in Eqs.~(\ref{eq:gap_nsym}) and (\ref{eq:defrho}) can be reduced to the standard form,
\begin{subequations}
\begin{eqnarray}
\Delta({\bf r},{\bf r}') = U(\tilde{r}) \sum _{E_{\nu}\ge 0}u_{\nu}({\bf r})v^{\ast}_{\nu}({\bf r}')[2f_{\nu}-1], 
\end{eqnarray}
\begin{eqnarray}
\rho _{\sigma}({\bf r}) = \sum _{E_{\nu} \ge 0}
\left[ 
|u_{\nu}({\bf r})|^2 f_{\nu} + |v_{\nu}({\bf r})|^2 (1-f_{\nu})
\right].
\end{eqnarray}
\end{subequations}
Hereafter, we use the definition in the case of a system without the time-reversal symmetry.

At low temperatures, the collisions between atoms can be described in terms of $s$-wave scattering, $U(\tilde{r}) \!=\! g \delta(\tilde{\bf r})$. Here, $g\!=\!4\pi^2 a/M$ is the coupling constant, and in general the interaction can be expressed using the dimensionless parameter $k_{\rm F}a$, where $k_{\rm F} \!\equiv\! \sqrt{2ME_{\rm F}}$ is the Fermi wave vector of a noninteracting Fermi gas with estimated Fermi energy $E_{\rm F}$. Then, the BdG equation (\ref{eq:bdg2}) is reduced to the following local form:
\begin{eqnarray}
\left[
\begin{array}{cc}
\mathcal{K}_{\uparrow}({\bf r}) & \Delta({\bf r}) \\
\Delta^{\ast}({\bf r}) & -\mathcal{K}^{\ast}_{\downarrow}({\bf r})
\end{array}
\right]
\left[
\begin{array}{c} u_{\nu}({\bf r}) \\ v_{\nu}({\bf r}) \end{array}
\right] = E_{\nu}
\left[
\begin{array}{c} u_{\nu}({\bf r}) \\ v_{\nu}({\bf r}) \end{array}
\right] ,
\label{eq:bdg_final}
\end{eqnarray}
where $\Delta({\bf r},{\bf r}')\!=\!\delta(\tilde{\bf r})\Delta({\bf r})$ and $\mathcal{K}_{\sigma}({\bf r},{\bf r}')\!=\!\delta(\tilde{\bf r})\mathcal{K}_{\sigma}({\bf r})$. 

Here, there are two singular contributions to the BdG equation (\ref{eq:bdg_final}) and gap equation (\ref{eq:gap_nsym}). First, the Hartree potential $g\rho _{\sigma}$ diverges at the unitary limit, which is neglected throughout this paper (see the text in \S~2.1, for details). The second singular behavior arises from the fact that the contact interaction leads to the UV divergence of the pairing field defined in eq.~(\ref{eq:gap_nsym}), where the leading term of $\Delta ({\bf r},{\bf r}')$ behaves as $-M\Delta({\bf r})/4\pi\tilde{r}$ at $\tilde{r}\rightarrow 0$:
\begin{eqnarray}
\Delta ({\bf r},{\bf r}') = -\frac{M}{4\pi\tilde{r}}\Delta({\bf R}) + g\mathcal{F}_{\rm reg} ({\bf R},\tilde{\bf r})
+ O(\tilde{\bf r}) ,
\label{eq:gapdiv}
\end{eqnarray}
where $\mathcal{F}_{\rm reg}$ is the regular part of the anomalous average $
\mathcal{F} ({\bf R},\tilde{\bf r}) \!=\! \langle {\psi}_{\downarrow}({\bf r}'){\psi}_{\uparrow}({\bf r})\rangle$
(${\bf R} \!\equiv\! ({\bf r}+{\bf r}')/2$ is the center-of-mass coordinate). One way to remove the divergent term is to replace the original contact interaction $g$ with the pseudopotential \cite{huang87, bruun, bulgac, grasso} and then, the formal expression for the pairing field is given by
\begin{eqnarray}
\Delta ({\bf r}) = g \lim _{\tilde{\bf r}\rightarrow 0} \frac{\partial}{\partial \tilde{r}} 
\left[ \tilde{r} \left\langle {\psi}_{\downarrow}({\bf r}'){\psi}_{\uparrow}({\bf r})\right\rangle \right] .
\label{eq:delta_formal}
\end{eqnarray}
To give a straightforward expression for the regularization operator $\lim _{\tilde{\bf r}\rightarrow 0}\partial _{\tilde{r}}[\cdot]$, we introduce the single-particle Green's function, $G_{\mu}({\bf r},{\bf r}') \!=\! \langle {\bf r} | H^{-1}_0 | {\bf r}' \rangle $ with $\delta\mu \!=\! 0$. This function yields the same nature of divergence as $\Delta$ in the limit of $\tilde{r} \!\rightarrow\! 0$: $ G_{\mu}({\bf R},\tilde{\bf r}) \!=\! \frac{M}{2\pi\tilde{r}} + G^{\rm reg}_{\mu}({\bf R}) + \mathcal{O}(\tilde{r})$. Here, the divergent contribution of the anomalous average in the right-hand side of eq.~(\ref{eq:delta_formal}) is canceled out by the irregular part of the single-particle Green's function, which allows one to introduce an explicit energy cutoff. By employing the LDA, we finally obtain the regularized gap equation \cite{bulgac,grasso},
\begin{eqnarray}
\Delta({\bf r}) = \tilde{g}({\bf r}) \sum _{\nu} u_{\nu}({\bf r}) v^{\ast}_{\nu}({\bf r})f_{\nu},
\label{eq:gap_final}
\end{eqnarray}
where the renormalized coupling constant $\tilde{g}({\bf r})$ is given by
\begin{eqnarray}
\frac{1}{\tilde{g}({\bf r})} = 
\frac{1}{g}+\frac{Mk_{\rm c}({\bf r})}{2\pi^2} \left[ 1 - \frac{k_{\rm F}({\bf r})}{2k_{\rm c}({\bf r})} 
\ln{\frac{k_{\rm c}({\bf r})+k_{F}({\bf r})}{k_{\rm c}({\bf r})-k_{\rm F}({\bf r})} }\right].
\label{eq:geff}
\end{eqnarray}
Here, $k_{\rm F}({\bf r})$ and $k_{\rm c}({\bf r})$ are the local wave vectors defined by the local Fermi and cutoff energies, respectively (see eq.~(\ref{eq:localenergy})).

\section{Matsubara Green's functions}

The Matsubara Green's function in the Nambu space is defined using the imaginary time $\tau$ as
\begin{eqnarray}
\hat{\mathcal{G}}(1,2) = -
\left\langle T_{\tau}\left[ 
\mbox{\boldmath $\Psi$}(1) \mbox{\boldmath $\Psi$}^{\dag}(2) 
\right] \right\rangle.
\end{eqnarray}
Here, we introduce the notation $1 \!\equiv\! ({\bf r}_1,\tau _1)$.
The Green's function in a $2\!\times\!2$ matrix form obeys the Gor'kov equation \cite{AGD}
\begin{eqnarray}
\int d2 \left[ - \frac{\partial}{\partial \tau}\hat{\tau}_0 \delta(1,2) 
- \hat{\mathcal{K}}(1,2) \right]\hat{\mathcal{G}}(2,1')
= \hat{\tau}_0 \delta (1,1'),
\label{eq:gorkov}
\end{eqnarray}
for a $2\!\times\! 2$ unit matrix $\hat{\tau}_0$. Also, we set $\hat{\mathcal{K}}(1,2) \!\equiv\! \delta(\tau _1 - \tau _2) \hat{\mathcal{K}}({\bf r}_1, {\bf r}_2)$. From the BdG equation (\ref{eq:bdg}), $\hat{\mathcal{K}}({\bf r}_1,{\bf r}_2)$ may be expanded as 
\begin{eqnarray}
\hat{\mathcal{K}}({\bf r}_1,{\bf r}_2) = \sum _{\nu} \hat{u}_{\nu}({\bf r}_1) \left[\begin{array}{cc}
\varepsilon ^{(\uparrow)}_{\nu} & 0 \\ 0 & - \varepsilon^{(\downarrow)}_{\nu}
\end{array} \right]
\hat{u}^{\dag}_{\nu}({\bf r}_2) .
\label{eq:}
\end{eqnarray}

The Green's function becomes diagonal in the representation that $\hat{\mathcal{K}}$ is diagonal. Hence, $\hat{\mathcal{G}}(1,2)$ may be expanded as
\begin{eqnarray}
\hat{\mathcal{G}}(1,2) = \frac{1}{\beta}\sum _{n}{\rm e}^{-i\omega _n (\tau _1 - \tau _2)}
\sum _{\nu} \hat{u}_{\nu}({\bf r}_1) \hat{\mathcal{G}}_{\nu}(i\omega _n)\hat{u}^{\dag}_{\nu}({\bf r}_2). \nn \\
\label{eq:green}
\end{eqnarray}
Its Fourier component $\hat{G}^{<(0)}(\varepsilon) $ is obtained easily as
\begin{eqnarray}
\hat{\mathcal{G}}_{\nu}(i\omega _n) = 
\left[ 
\begin{array}{cc}
\displaystyle{\left(i\omega _n - \varepsilon^{(\uparrow)}_{\nu}\right)^{-1}} & 0 \\ 0 & 
\displaystyle{\left(i\omega _n + \varepsilon^{(\downarrow)}_{\nu}\right)^{-1}}
\end{array}
\right]. \label{eq:g0}
\end{eqnarray}

It is convenient to introduce the Fourier transform with respect to $\tau$, $\hat{\mathcal{G}}(1,2) \!=\! \beta^{-1}\sum _{n} {\rm e}^{-i\omega _n (\tau _1 - \tau _2)}\hat{\mathcal{G}}({\bf r}_1{\bf r}_2,i\omega _n)$, whose coefficient is obtained from eq.~(\ref{eq:green}). From the eigenfunctions and energy of the BdG equation, $(u_{\nu},v_{\nu})$ and $E_{\nu}$, $\hat{\mathcal{G}}({\bf r}_1{\bf r}_2,i\omega _n)$ is given by 
\begin{subequations}
\label{eq:green2}
\begin{eqnarray}
\mathcal{G}_{11}({\bf r}_1{\bf r}_2, i\omega _n) 
= \sum _{\nu}\frac{u_{\nu}({\bf r}_1)u^{\ast}_{\nu}({\bf r}_2)}{i\omega _n - E_{\nu}}, 
\end{eqnarray}
\begin{eqnarray}
\mathcal{G}_{12}({\bf r}_1{\bf r}_2, i\omega _n) 
= \sum _{\nu}\frac{u_{\nu}({\bf r}_1)v^{\ast}_{\nu}({\bf r}_2)}{i\omega _n - E_{\nu}}, 
\end{eqnarray}
\begin{eqnarray}
\mathcal{G}_{21}({\bf r}_1{\bf r}_2, i\omega _n) 
= \sum _{\nu}\frac{v_{\nu}({\bf r}_1)u^{\ast}_{\nu}({\bf r}_2)}{i\omega _n - E_{\nu}}, 
\end{eqnarray}
\begin{eqnarray}
\mathcal{G}_{22}({\bf r}_1{\bf r}_2, i\omega _n)  
= \sum _{\nu}\frac{v_{\nu}({\bf r}_1)v^{\ast}_{\nu}({\bf r}_2)}{i\omega _n - E_{\nu}},
\end{eqnarray}
\end{subequations}
where $\mathcal{G}_{ij}$ denotes the $(i,j)$ element of the $2\!\times\! 2$ matrix $\hat{\mathcal{G}}$.

\end{document}